\documentclass[10pt,prd]{revtex4}

\usepackage{amssymb}
\usepackage{amsmath}
\usepackage{graphicx}
\usepackage{dcolumn}
\usepackage{bm}
\usepackage{hyperref}
\usepackage{physics}
\usepackage{mathtools}
\usepackage{subcaption}

\allowdisplaybreaks

\usepackage{graphicx,amssymb,amsmath,amsthm,amsfonts}
\usepackage[usenames]{color}
\usepackage{xcolor}
\usepackage{mathtools}

\begin{document}

\title{Black holes and hot shells in the Euclidean path integral
approach to quantum gravity}
\author{Jos\'{e} P. S. Lemos}
\affiliation{Center for Astrophysics and Gravitation -
CENTRA, Departamento de F\'{\i}sica,
Instituto Superior T\'{e}cnico - IST, Universidade de Lisboa - UL,
Avenida Rovisco Pais 1, 1049-001, Portugal;
email: joselemos@ist.utl.pt.}
\author{O. B. Zaslavskii}
\affiliation{Department of Physics and Technology,
Kharkov V.~N.~Karazin National
University, 4 Svoboda Square, Kharkov 61022, Ukraine;
email: zaslav@ukr.net}

\begin{abstract}
We study a spherical black hole surrounded by a hot self-gravitating
thin shell in the canonical ensemble, i.e., a black hole and a hot
thin shell inside a heat reservoir acting as a boundary with its area
and temperature fixed. To work out the quantum statistical mechanics
partition function, from which the thermodynamics of the system
follows, we use the Euclidean path integral approach to quantum
gravity that identifies the path integral of the gravitational system
with the partition function itself. In a semiclassical evaluation of
the path integral, one needs to compute the classical action of the
system.  From the action, one finds the result that the total entropy,
given by the sum of black hole and matter entropies, is a function of
the gravitational radius of the system alone. So, the black hole
inside the shell has no direct influence on the total entropy.  One
also finds the free energy which is equal to the action times the
temperature, the thermodynamic energy, and the temperature
stratification along the system.  Another important result is that the
heat reservoir temperature is composed of a free function of the
gravitational radius of the system,
which acts as a reduced temperature equation
of state, divided by the redshift function at the reservoir.  Upon the
specification of the reduced temperature, the solutions for the
gravitational radii of the system compatible with the boundary data
can be found.  In addition, it is found that the black hole inside the
shell has two possible horizon radii. The first law of thermodynamics
is then identified, and it is shown that there is a first law for the
system, a first law for the hot shell, and yet another first law for
the black hole.  The thermodynamic stability analysis is performed
through the calculation of the system's heat capacity.  By specifying
for the free function the Hawking temperature equation of state for
the gravitational radius of the system, which is not a black hole, one
finds a remarkable exact mechanical and thermodynamic solution. With
the exact solution in hand one establishes that pure black hole
spaces, hot shell with a black hole spaces, pure hot shell spaces, and
hot flat spaces are phases that cohabit in the ensemble, with some of
them acting as thermodynamic mimickers.  This exact solution for a
black hole with a self gravitating hot shell is not only of interest
in itself, but can also be seen as a model to situations involving
black holes interacting with hot gravitons and other hot particles.
The study of the high temperature limits for the system also reveals
several important aspects.
\end{abstract}

\keywords{Path integral, black holes, hot matter, thermodynamics,
phase transitions}
\maketitle

\section{Introduction}

The path integral formulation of quantum mechanics is well adapted to
describe quantum relativistic systems including the ones where
gravitation is important.  Applying this method to particles on a
black hole background, Hartle and Hawking \cite{hh1976} were able to
derive quite naturally the Hawking radiance temperature
$T_{\rm H}=\frac{1}{4\pi r_+}$, with $r_+$ being the gravitational radius of
the black hole.

Moreover, when time is Euclideanized, the corresponding Euclidean path
integral formulation yields directly the quantum partition function of
a system in the statistical mechanics canonical ensemble, from which
the thermodynamics of the system itself can be derived.  The
application of statistical mechanics and thermodynamics to gravitation
and black holes is of fundamental importance, since on one hand
statistical mechanics and thermodynamics give clues to what are the
degrees of freedom of a system, on the other hand a black hole is the
object par excellence to bring together gravitation and quantum
mechanics. Statistical mechanics makes use of the concept of
ensembles, where collections of thermodynamic reservoirs are analyzed
statistically. There are several different types of ensembles, the
one that is relevant to the Euclidean path integral
is the canonical ensemble in which the reservoirs are kept at
a fixed temperature and a fixed size.  The recognizing of the
potentiality of the path integral when applied to black hole physics,
led to apply it to quantum gravity in general. An initial endeavor
undertook by Gibbons and Hawking \cite{GibbHawk1977} employed the
Euclidean path integral to quantum gravity to understand black hole
thermodynamics through the Euclideanization of the time coordinate of
the corresponding Euclidean black hole solution. There would crop up,
however, a problem. The system,
with the Hawking temperature fixed at infinity,
would be unstable, the black hole would
radiate gravitons or other matter fields on and on, equilibrium could
never be achieved, the canonical ensemble would not exist, and
ultimately the approach would not be well formulated. Further work on
the instability of such a black hole through first order, or one-loop,
perturbations was elaborated by Gross, Perry, and Yaffe \cite{gpy}.
Hawking and Page \cite{hawkingpage} managed
to suppress this canonical ensemble instability by embedding
the black hole in a Euclidean anti-de Sitter space which has a natural
effective boundary at infinity.  A further advancement was made by Allen
\cite{allen} that found that the Euclidean Schwarzschild solution
could be stabilized if the
gravitational perturbations 
were performed with a fixed temperature boundary condition at
some finite
radius.

York \cite{york1} then understood that the
Euclidean path integral approach to quantum asymptotically flat black
holes is well posed only when proper data at a compact boundary is
given.  
The corresponding canonical ensemble is then comprised of systems of
heat reservoirs with a fixed temperature and a fixed area, each
containing a black hole. When the classical general relativistic
action is evaluated for the system, it turns out that it gives a
contribution to the path integral already at zeroth order, and thus
the quantum statistical mechanics partition function at this order is
given by the exponential of the classical Euclidean action.  The
thermodynamics of the system follows, since the logarithm of the
partition function is essentially the free energy 
divided by the temperature, i.e., the free energy can be identified to
the classical action times temperature of the system.  Remarkably,
compatible with the reservoir boundary data there are two black holes,
one small, when compared with the reservoir size, which is unstable
and corresponds to the Gibbons-Hawking black hole,
and the other large
which is stable. The method was generalized by Whiting and York
\cite{whitingyork} to include other possible formal spherical vacuum
solutions with some quantum features, and then applied to include
matter in the form of thin shells by Martinez and York
\cite{ym}. Several developments and applications were done using the
Euclidean path integral applied to black hole systems in the canonical
ensemble following York's formalism.
Hayward \cite{hayward0} implemented the approach to spaces with a
de Sitter cosmological horizon, Zaslavskii \cite{can} analyzed generic
static configurations without assumption of spherical symmetry,
Hayward \cite{hayward} studied the first law of thermodynamics of
systems with matter and black hole and cosmological horizons, and
Zaslavskii treated the thermodynamics of black holes with skyrme
matter fields \cite{zalavskiionskyrme}.
The three-dimensional anti de Sitter black hole in the canonical
ensemble was investigated by Zaslavskii \cite{zaslavskiibtz}, as well
as by Brown, Greighton, and Mann \cite{browcreima} that also studied
the four-dimensional Schwarzschild-anti de Sitter black hole.
Yu \cite{hongwei} considered the Euclidean action and thermodynamics
of a black hole with a global monopole, and Lemos \cite{lemos1996}
examined the two-dimensional black hole that appears in the
Teitelboim-Jackiw theory.
Further analysis of the Schwarzschild-anti de Sitter black hole
including its generalization to incorporate electric charge in a larger
ensemble was performed by Pe\c ca and Lemos \cite{lemospeca}.  Gregory
and Ross \cite{gregoryross} showed that the negative modes found
previously for first order perturbations in the gravitational action
of the Euclidean Scharzschild black hole in the canonical ensemble are
indeed strictly correlated with the thermodynamic instability of those
black holes,
Lemos and Zaslavskii \cite{lz1} applied the
Euclidean action approach to the 
membrane paradigm to find
the black hole entropy,
Zaslavskii scrutinized distorted vacuum black holes in
the canonical ensemble \cite{zasla2019}, Andr\'e and Lemos \cite{la1}
considered Schwarzschild-Tangherlini black holes in five dimensions in
the canonical ensemble, where there are exact solutions for the small
and unstable and the large and stable black holes,and then generalized
the formalism to the $d$-dimensions \cite{la2}.  In canonical ensemble
systems, the thermodynamics of gravitational and gravitoscalar fields
in black hole and cosmological horizon backgrounds was exhibited by
Miyashita \cite{miyashita}, cosmological de Sitter horizons were
further investigated by Banihashemi and Jacobson
\cite{banihashemijacobson}, and causal regions in several settings,
including the two-dimensional Jackiw-Teitelboim theory, were discussed
by Jacobson and Visser \cite{jacobsonvisser}

The thermodynamic analysis of gravitational systems with matter can
proceed in different ways. For instance, Martinez \cite{mart} analyzed
the thermodynamics of thin shells with Minkowski spacetime inside and
Schwarzschild outside by investigating the possibilities of the first
law of thermodynamics on the system, a work that was generalized by
Andr\'e and Lemos \cite{jr} to $d$-dimensional shells with a
Schwarzschild-Tangherlini outside spacetime, and that was further
studied by Perez, Chiapparini, and Reyes \cite{berg}
and by Fernandes and Lemos \cite{fernandeslemos}.
Other
approaches to gravitational matter systems use the quasiblack holes
of Lemos and Zaslavskii,
which can be identified as black hole mimickers \cite{lzmimickers},
and yield a precise manner to calculate the black hole
entropy
\cite{qbhent}, its mass formula
displaying the analogy of the membrane paradigm
with the quasiblack hole
approach
\cite{lzmassf}, and
analyzing several other properties
within the approach
\cite{lz3}.

Our aim is to understand more deeply the quantum and thermodynamic
properties of microscopic gravitational systems involving black holes
and hot matter. Since a pure black hole loses mass to infinity in the
form of gravitons or other matter fields through the Hawking emission
process until it eventually disappears, in order to determine the
interplay between pure black holes states on one hand and matter
fields in curved space states on the other hand, and look for the
possible phase transitions between those states, one encloses them
inside a heat reservoir at constant temperature and constant radius.
To find the statistical mechanics partition function at a
semiclassical level for this canonical ensemble we use the Euclidean
path integral approach to quantum gravity.  From the partition
function one then is able to obtain all the necessary quantities
required to make a full thermodynamic analysis.  We model the hot
matter fields by a hot matter
self gravitating thin shell, and so, specifically, our
microscopic gravitational system is composed of a heat reservoir
surrounding a static spherical hot matter thin shell that harbors a
black hole in its interior.  We present several important results. One
is that the thermodynamics is controlled 
by the gravitational
radius of the system.
Another is that, upon giving a specific equation of state for
the temperature that yields an exact solution, several thermodynamic
phases, like pure black holes, hot shell spaces with a black hole
inside, hot shell spaces with nothing inside, and hot flat spaces,
that can cohabit in the ensemble, show up.  Yet another result
connected with the conversion of pure black holes into curved spaces
with hot matter in the form of hot thin shells and vice versa, is that
the systems in some instances perform as black hole thermodynamic
mimickers as well as dynamic and geometric mimickers. There are more
results of interest concerning these microscopic gravitational systems
that can be treated semiclassically, composed of hot matter thin shells
and black holes with quantum properties that will be presented along
the text.

The paper is organized as follows.  In Sec.~\ref{mechanics}, we
display the spacetime and the mechanical structure of a static
spherical thin shell with a black hole inside.  In
Sec.~\ref{canonical}, we define the canonical ensemble of the system,
give the action for the system which in turn yields directly in the
semiclassical approximation the corresponding quantum partition
function, and do a thorough analysis on the entropy showing that it
depends on the gravitational radius of the system alone.  In
Sec.~\ref{canonicalenergy} we
give the expression for the free energy,
perform a thermal energy analysis, and do a
complete temperature analysis for the reservoir, the hot thin shell,
and the black hole.  In
Sec.~\ref{firstlaw} we show that the first law of thermodynamics
applies to the three relevant cases, namely, the entire system, the
hot thin shell, and the black hole, we clarify the consistency of the
whole scheme, and we perform a
thermodynamic stability analysis for the system by
calculating its heat capacity.  In Sec.~\ref{specificeos} we present a
specific temperature equation of state for the system, specifically
the Hawking temperature equation of state, find the action and
entropy, find the thermodynamic exact solutions, namely, the solutions
for the exterior gravitational radius of the hot thin shell and the
solutions for the interior horizon radius of the black hole, study
their stability, and work out all the phases of the semiclassical full
spectrum of solutions of the system, namely, black holes, hot shell
spaces, and hot flat space. 
In Sec.~\ref{highT} we perform the high temperature limits of the
shell and black hole system in the heat reservoir, showing that there
are four cases depending on the limiting values one takes for the
gravitational radius of the hot shell and for the horizon radius of
the black hole.  In Sec.~\ref{conc} we conclude.
In the Appendix~\ref{othereos} we comment on the
results that might arise for generic temperature
equations of state,
with emphasis on which semiclassical phases may dominate the ensemble.

We use units such
that the constant of gravitation, the velocity of light, the Planck
constant, and the Boltzmann constant are all set to one, so that all
the formulas and expressions are quantum expressions in a
semiclassical approximation.

\section{Black hole surrounded by a
self-gravitating thin shell: Spacetime and
mechanics}
\label{mechanics}

We analyze here the general relativistic spacetime features and the
mechanics of a system consisting of a black hole surrounded by a
self gravitating thin
matter shell, the whole system being static and spherically symmetric.

The spacetime is divided into three regions, one region is the inside
of the shell containing a black hole, the other region is the shell
itself, and the third region is the region outside the shell.  The
interior region has a Schwarzschild black hole metric form, the shell
has radius $\alpha$, and the exterior region to the shell has a
Schwarzschild metric form with parameters that are different from the
parameters of interior metric.  Then, the line elements for the
interior and exterior regions are
\begin{align}\label{met1}
ds^2=- &\left(1-\frac{r_+}{r}\right)
\left(\frac{1-\frac{\tilde{r}_+}{\alpha }}{1-
\frac{r_+}{\alpha }}\right)
dt^2 +
\frac{dr^2}{1-\frac{r_+}{r}} +
r^2\left(d\theta^2+\sin^2\theta\,d\phi^2\right)\,,\quad
 r_+<r < \alpha\,,
\end{align}
\begin{align}\label{met2}
ds^2=-\left(1-\frac{\tilde{r}_+}{r}\right)dt^2+
\frac{dr^2}{1-\frac{\tilde{ r}_{+}}{r}} +
r^2\left(d\theta^2+\sin^2\theta\,d\phi^2\right)\,,\quad\quad
\quad\quad \quad\quad 
\alpha < r <\infty\,, 
\end{align}
respectively, with $-\infty<t<\infty$,
$0\leq\theta\leq\pi$, $0\leq\phi<2\pi$, and 
where $r_+$ is the event horizon radius of the black hole and
$\tilde{r}_+$ is the gravitational radius of the thin shell,
$\tilde{r}_+$ not being an event horizon radius.  Considering that the
black hole has mass $M$, then $r_+=2M$, and
considering that the outer spacetime
has mass $\tilde{ M}$, then $\tilde{r}_+=2\tilde{M}$.
Note that $r_+\leq \tilde{r}_+\leq\alpha$.
To define the
line element at $\alpha$, which is not defined neither in
Eq.~\eqref{met1} nor in~\eqref{met2}, one has to be careful, but
we do not
need to go into the details here.  In Eqs.~\eqref{met1}
and~\eqref{met2} one can extract three important functions.  For the
interior, from Eq.~\eqref{met1}, one defines the function $k(r)$ as
\begin{equation}
k(r)= \sqrt{1-
\frac{r_+}{r}}\,,\quad r_+\leq r \leq \alpha
\,.
\label{reds1}
\end{equation}
Note that $k(r)$ is in fact
$k(r)=k(r,r_+)$.
For the exterior, from Eq.~\eqref{met2}, one defines the function
$\tilde{k}(r)$ as
\begin{equation}
\tilde{k}(r)=\sqrt{1- \frac{\tilde{r}_+}{r}}\,,\quad \alpha
\leq r <\infty\,.
\label{redstilder}
\end{equation}
Note also that $\tilde{k}(r)$ is in fact
$\tilde{k}(r)=\tilde{k}(r,\tilde{r}_+)$.
At the shell,  from Eq.~\eqref{met1},
there is the junction function $B(\alpha)$
defined by 
\begin{equation}
B(\alpha)=\frac{\tilde{k}(\alpha)}{{k}(\alpha)}\,,\quad r=\alpha\,,
\label{completesreds1}
\end{equation}
where
from Eqs.~\eqref{reds1} and~\eqref{redstilder}, one has
$k(\alpha)=\sqrt{1- \frac{r_+}{\alpha}}$ and
$\tilde{k}(\alpha)=\sqrt{1- \frac{\tilde{r}_+}{ \alpha}}$,
respectively.
Of course,
$k(\alpha)=k(\alpha,r_+)$, 
$\tilde{k}(\alpha)=\tilde{k}(\alpha,\tilde{r}_+)$,  and
$B(\alpha)=B(\alpha,\tilde{r}_+,r_+)$.
The redshift in the interior region is given by the function
$k(r)B(\alpha)$, which can be taken from
Eqs.~\eqref{reds1} and~\eqref{completesreds1},
and the redshift in the exterior region is given by the function
$\tilde{k}(r)$, which can be taken from
Eq.~\eqref{redstilder}.

Now in 
making the
match
between the interior and exterior spacetimes,
Eqs.~\eqref{met1}
and \eqref{met2},
there are two junction conditions
that must be satisfied at $\alpha$ and give the matter properties
of the shell. 
One junction condition
gives the surface energy density $\sigma_{\rm m}$
of the matter
at the shell as
\begin{equation}
\sigma_{\rm m}=\frac{1}{4\pi\alpha}
\left(k(\alpha)-\tilde{k}(\alpha)\right)\,.
\label{junctionmass}
\end{equation}
The  surface energy density $\sigma_{\rm m}$
at $\alpha$, is thus a function of $\alpha$,
$\sigma_{\rm m}=\sigma_{\rm m}(\alpha)$, as well as a function of
$\tilde{r}_+$ and $r_+$, so in
full $\sigma_{\rm m}=\sigma_{\rm m}(\alpha,\tilde{r}_+,r_+)$.
This can be explicitly inverted to give
$\tilde{r}_+=\tilde{r}_+(r_+,\sigma_{\rm m},\alpha)$.
The other junction condition gives the
tangential pressure $p_{\rm m}$ at the shell, which can be calculated to be 
$8\pi p_{\rm m}=\frac{\tilde{k}(\alpha)-k(\alpha)}{\alpha }+
(\tilde{k}^{\prime}(\alpha)
-{k}^{\prime}(\alpha))$, i.e., 
\begin{equation}
p_{\rm m}=
\frac1{16\pi\alpha}
\frac{
\left({k}(\alpha)-\tilde{k}(\alpha)\right)
\left(1-{k}(\alpha)\,\tilde{k}(\alpha)\right)}
{{k}(\alpha)\,\tilde{k}(\alpha)}\,.
\label{pressshell1}
\end{equation}
One has
that $p_{\rm m}=p_{\rm m}(\alpha)$,
or in full $p_{\rm m}=p_{\rm m}(\alpha,\tilde{r}_+,r_+)$.
The spacetime and its mechanics have now been characterized.
In the particular case that there is no black hole
in the interior, one has $r_+=0$, the inside of the shell
is flat spacetime with the outside of the shell being
Schwarzschild with gravitational radius 
$\tilde{r}_+$.

\section{Black hole surrounded by a hot
self-gravitating thin shell and a heat
reservoir in the canonical ensemble: Euclidean action and total
entropy of the system} \label{canonical}

\subsection{The  canonical ensemble,
the heat reservoir, and the Euclidean action}

We now want to place the
self-gravitating shell and black hole inside a heat reservoir
and treat the problem as a
canonical ensemble problem in statistical mechanics.
The canonical ensemble is realized by a collection of equal
systems for which the size and temperature are kept constant. To
keep a system at constant temperature one encloses it in a very large
heat bath reservoir with the property
that the system and the heat bath can exchange energy.

One way to proceed is to use the Euclidean path integral approach
which is well adapted to study statistical mechanics canonical
ensembles and that can be applied to quantum gravity systems too.
Denoting the gravitational field by $g$ and the possible other matter
fields by $\phi$ the quantum partition function for the system can be
written as $Z= \int d[g,\phi] {\rm e}^{-I[g,\phi]} $, where the
integral is taken over all gravitational fields $g$ and other fields
$\phi$ which are periodic with period $\beta$ in imaginary time. $Z$
is the function suited to study systems at a given temperature $T$,
after the identification $T=\frac1\beta$. From the partition function
one obtains the thermodynamic free energy $F$ of the system, given
through the relation $\beta F=-\ln Z$, i.e.,
$F=-T\ln Z$. The other thermodynamic
quantities such as the entropy $S$ and the thermodynamic energy $E$
then follow.

Once the full action $I$ for the system is known, and upon giving the
boundary conditions for the system, which in the canonical ensemble
means giving the temperature and the size of the system, the partition
function $Z$ can then be formally calculated with full precision.  In
practice, one has to resort to approximations.  For gravitational
systems, one can choose a zero order, or semiclassical, approximation,
in which the part of the action that most contributes is the classical
action. In this case the integral for calculating $Z$ drops and one
gets $Z= {\rm e}^{-I}$, where $I$ now refers to the classical action
of the system.  Note that although $Z$ is dictated in this
approximation by the classical action, $Z$ evanesces in the classical
limit, i.e., it vanishes when the Planck constant goes to zero, and so
the partition function $Z$ is indeed a quantum gravitational quantity.
From $\beta F=\ln Z$, one deduces that in the semiclassical
approximation one has $I=\beta F$, i.e., $F=TI$. Then, from $I$ or $F$
for the gravitational field plus matter system one obtains all
the relevant thermodynamics in this order of approximation.

In the semiclassical approximation, the Euclidean total action $I$ in
a heat reservoir problem is a sum of the Euclidean gravitational
action $I_{\rm g}$ and a Euclidean matter action $I_{\rm m}$, i.e.,
$I=I_{\rm g}+I_{\rm m} $.  The Euclidean gravitational action
appropriate when there is a fixed boundary, in this case the
reservoir, is given by $I_{\rm g}=-\frac{1}{16\pi}\int_{\cal M} {\cal
R}\, \sqrt{g}\, d^4x+ \frac{1}{8\pi}\int_{\cal \partial M} ({\cal
K}-{\cal K}_0)\, \sqrt{\gamma}\,d^3x $, where the first integral is
the bulk part performed over the four-dimensional manifold ${\cal M}$,
with ${\cal R}$ being the Ricci scalar and $g$ being the determinant
of the Euclidean metric, and the second integral is the boundary term
part performed over the three-dimensional boundary ${\partial \cal
M}$, with ${\cal K}$ being the extrinsic curvature of the boundary,
${\cal K}_0$ being the extrinsic curvature of the reference Euclidean
Minkowski space, and $\gamma$ being the determinant of the induced
boundary metric.  The Euclidean matter action applicable to our
purposes is given by $I_{\rm m}=\int_{\cal M} {\cal L}_{\rm m}
\sqrt{g}\,d^4x $, where the integral is performed over the manifold
${\cal M}$, $g$ is the determinant of the Euclidean metric, and ${\cal
L}_{\rm m}$ is the matter Lagrangian that one chooses to use in a
canonical ensemble system. For instance, one can put ${\cal L}_{\rm
m}=\rho_{\rm m}-T_{\rm m}s_{\rm m}$, with $\rho_{\rm m}$ being the
energy density of the matter, $T_{\rm m}$ its temperature, and $s_{\rm
m}$ its entropy density, i.e., the matter Lagrangian is the one
appropriate for hot matter having as thermodynamic potential the
thermodynamic canonic potential.

\begin{figure}[h]
\begin{center}
\includegraphics*[scale=1.0]{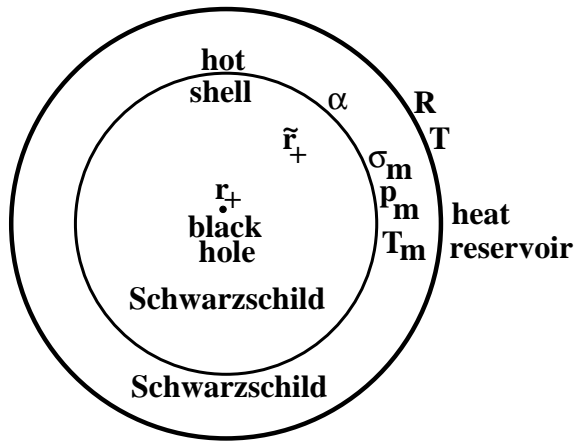}
\end{center}
\caption{A drawing of a black hole within a hot
self-gravitating thin shell inside a heat
reservoir at temperature $T$ and radius $R$.  Outside the black hole
the geometry is a Schwarzschild one, and outside the shell there is
another Schwarzschild geometry with different parameters.  The
Euclideanized space and its boundary have $R^2 \times S^2$ and
$S^1 \times S^2$ topologies, respectively, where the
$S^1$ subspace with proper length $\beta=\frac1T$ is not displayed.
See text for more details.}
\label{bhplusthinshellcanonical}
\end{figure}

The canonical ensemble we want to study is realized by a hot
self-gravitating thin
shell which surrounds a black hole and is enclosed in a heat
reservoir, see Fig.~\ref{bhplusthinshellcanonical}.  To advance, one
Euclideanizes time by putting $\tau=-it$ and gets the Euclidean line
elements of Eqs.~\eqref{met1} and \eqref{met2} for the interior and
exterior space, respectively. Then, one puts a boundary $S^1 \times
S^2$ to the Euclideanized space with $R^2 \times S^2$ topology, where
the circumference $S^1$ has proper length $\beta$ with
\begin{equation}
\beta=\frac1T
\,,
\label{inverseT}
\end{equation}
$T$ being the temperature at the boundary, and where
the boundary
sphere $S^2$ has area $A$ equal to
\begin{equation}
A=4\pi R^2\,,
\label{areacavity}
\end{equation}
with $R$ being the rdius defined by
the area $A$.  In order that the
Euclideanized manifold does not have conical singularities in the
plane specified by the Euclidean time and the radial coordinate, and
so the action is not singular, one has to impose that the Euclidean
time $\beta$ has a well determined period.  The canonical ensemble for
the system is characterized by these two values at the boundary,
namely,
the inverse temperature $\beta$ or,
which is the same, the temperature $T$, and the radius
$R$. The black hole plus hot matter system that lives inside the
boundary has necessarily to adapt to these boundary values.

\vskip 0.6cm
\vskip 0.4cm

The Euclidean total action $I$ for the black hole and
hot self-gravitating 
thin shell in a
heat reservoir problem, namely, $I=I_{\rm g}+I_{\rm m}$, can be
usefully split into an action $I_{\rm in}$ for the inside which
contains a black hole, an action for the middle $I_{\rm shell}$ which
contains a thin matter shell, and an action for the outside $I_{\rm
out}$ that extends from the shell up to the reservoir boundary, i.e.,
$I=I_{\rm in}+I_{\rm shell}+I_{\rm out}$.  The $I_{\rm in}$ and
$I_{\rm out}$ actions have pure geometric terms. The shell action
includes a geometrical term and a matter term, where the energy
density $\rho_{\rm m}$
should here be identified as the surface energy density
$\sigma_{\rm m}$ of the matter, since $\rho_{\rm m}$ has support
only on the shell,
the temperature of the matter $T_{\rm m}$ is identified with the
temperature of the shell, $T_{\rm shell}$, which is also the
temperature at the shell $T(\alpha)$, so that $T_{\rm m}\equiv T_{\rm
shell}\equiv T(\alpha)$, and the entropy density $s_{\rm m}$
is identified
with the surface entropy density, also denoted
by $s_{\rm m}$. One can assume that there is an equation of
state for the matter on the shell in which the
energy density  $\sigma_{\rm m}$
depends on the entropy density $s_{\rm m}$
and on the size $\alpha$, i.e., 
$\sigma_{\rm m}=\sigma_{\rm m}(s_{\rm m},\alpha)$ or,
inverting it if one wishes,
an equation of state of the form
$s_{\rm m}=s_{\rm
m}(\sigma_{\rm m},\alpha)$.

To find the reduced  Euclidean action for the system
an important feature that has to be taken into account
in the calculation is that
the inverse temperature, is fixed at the boundary $R$.
This implies that, since
we are dealing with a Euclideanized time with the
circumference
$S^1$ at the
boundary with radius $R$ having length $\beta$, one can write
$\beta=P\tilde{k}(R)$ for an arbitrary fixed period $P$.
Moreover, the Euclidean
two-manifold
spanned by the periodic Euclidean time and radius
cannot have
conical singularities, and so 
one must impose
that
$\left[ {k}(r) {k}^\prime(r)
\dfrac{ \tilde{k}(\alpha)}
{{k}(\alpha)} \right]_{r_+}
=\dfrac{2\pi}{P}$.
A careful treatment of the Euclidean
gravitational plus matter action can then
be made yielding the following
reduced action
\begin{equation}
I=\beta R\left[1-\tilde{k}(R)\right]
-\pi r_+^2-S_{\rm m}
\,,
\label{action0}
\end{equation}
where $\tilde{k}(R)=\sqrt{1-\frac{\tilde{r}_+}{R}}$ is the redshift
factor at the boundary $R$, see Eq.~\eqref{redstilder}, and $S_{\rm
m}=4\pi \alpha^2
s_{\rm m}
$ is the entropy of the matter in the
thin shell, see also \cite{ym,hayward0,can,hayward}.
Note that 
in Eq.~\eqref{action0} the reduced
action $I$ can be envisaged
as $I=I(\beta,R;r_+,\alpha,s_{\rm m})$,
where $\beta$ and $R$ are quantities that are kept
constants, and $r_+$, $\alpha$, and $s_{\rm m}$
are quantities that can vary.
The dependence of the action on 
$r_+$, $\alpha$, and $s_{\rm m}$, can be
seen, since
$\tilde{r}_+=\tilde{r}_+(r_+,\alpha,\sigma_{\rm m})$, see
Eq.~\eqref{junctionmass},
and the matter can be assumed to have
an equation of state of the   form
$\sigma_{\rm m}=\sigma_{\rm m}(\alpha,s_{\rm m})$.
To find the minimal value of $I$, the one that
most contributes to the path integral and
thus to the partition function, one sets
$dI=0$. Since $dI=
\frac{\partial I}
{\partial r_+} dr_+
+
\frac{\partial I}
{\partial \alpha} d\alpha
+
\frac{\partial I}
{\partial s_{\rm m}} ds_{\rm m}
$, this means
that the stationary points
of the action are obtained
through the equations
$\frac{\partial I}
{\partial r_+}=0$,
$\frac{\partial I}
{\partial \alpha}=0$, and
$\frac{\partial I}
{\partial s_{\rm m}}=0$,
with $\beta$ and $R$ held fixed.
We follow another path.

\subsection{Entropy analysis for black hole and hot thin shell in the
canonical ensemble: Entropy dependence on the gravitational radius of
the system only, $S=S(\tilde{r}_+)$}

\subsubsection{Total entropy of the system}

Inverting Eq.~\eqref{junctionmass} one finds
$\tilde{r}_+= r_+ + 8\pi\alpha^2\sigma_{\rm m} \,k(\alpha)-
16\pi^2\alpha^3\sigma_{\rm m}^2$, i.e.,
$\tilde{r}_+=\tilde{r}_+(r_+,\alpha,s_{\rm m})$.
So, in the action $I=I(\beta,R;r_+,\alpha,s_{\rm m})$
of Eq.~\eqref{action0}, we
can 
swap variables
from $r_+,\alpha,s_{\rm m}$
to $\tilde{r}_+,\alpha, s_{\rm m}$,
to yield
$I=I(\beta,R;\tilde{r}_+,\alpha,s_{\rm m})$.
In this case, one has $dI=
\frac{\partial I}
{\partial \tilde{r}_+} d\tilde{r}_+
+
\frac{\partial I}
{\partial \alpha} d\alpha
+
\frac{\partial I}
{\partial s_{\rm m}} ds_{\rm m}
$, and
the minimal value of $I$,
obeying 
$dI=0$, yields 
that the stationary points
of the action are obtained
through the equations
$\frac{\partial I}
{\partial  \tilde{r}_+}=0$,
$\frac{\partial I}
{\partial \alpha}=0$, and
$\frac{\partial I}
{\partial s_{\rm m}}=0$.

Now, the first part of the action $I$ in Eq.~\eqref{action0} is the
term $\beta R\left[1-\tilde{k}(R)\right]$, so that keeping $\beta$ and
$R$ constants we have $\frac{\partial\left(\beta
R\left[1-\tilde{k}(R)\right]\right)} {\partial
\tilde{r}_+}=\frac{\beta}{2\,\tilde{k}(R)}$,
$\frac{\partial\left(\beta R\left[1-\tilde{k}(R)\right]\right)}
{\partial \alpha}=0$, and $\frac{\partial\left(\beta
R\left[1-\tilde{k}(R)\right]\right)} {\partial s_{\rm m}}=0$.  The
second part of the action $I$ in Eq.~\eqref{action0} contains the sum
of two terms, $\pi r_+^2+S_{\rm m}$.  Let us call this sum $S$, so that
\begin{equation}
S=\pi r_+^2+S_{\rm m}\,.
\label{Swithshellandbh}
\end{equation}
Its total differential
is $dS$,
with 
$dS=
\frac{\partial S}
{\partial \tilde{r}_+} d\tilde{r}_+
+
\frac{\partial S}
{\partial \alpha} d\alpha
+
\frac{\partial S}
{\partial s_{\rm m}} ds_{\rm m}
$.
So the condition $dI=0$
implies in the variables 
$\tilde{r}_+,\alpha,s_{\rm m}$
the following equation
$\frac{\partial\left(\beta R\left[1-\tilde{k}(R)\right]\right)}
{\partial  \tilde{r}_+}d\tilde{r}_+-dS=0$, i.e.,
$dS=\frac{\beta}{2\,\tilde{k}(R)}\,d\tilde{r}_+$.
Since $S$ has to be an exact differential
in the space of configurations in equilibrium,
the only way
that this can happen
is to set $\frac{\beta}{\,\tilde{k}(R)}=b(\tilde{r}_+)$,
so that,
\begin{equation}
\beta=b(\tilde{r}_+)\tilde{k}(R),
\label{tolman2R}
\end{equation}
where $b(\tilde{r}_+)$ is a well-behaved, but otherwise
arbitrary, function of 
$\tilde{r}_+$, and $\tilde{k}(R)$ is the
redshift function at $R$ due to gravitational
radius $\tilde{r}_+$, see Eq.~\eqref{redstilder}.
The function $b(\tilde{r}_+)$ is in its essence
an equation of state for the inverse temperature
of the system that must be supplied.
Thus, we find
$\frac{\partial S}
{\partial \tilde{r}_+} =\frac12
\frac{\beta}{\,\tilde{k}(R)}=\frac12 b(\tilde{r}_+)$,
$\frac{\partial S}
{\partial \alpha} =0$, and
$
\frac{\partial S}
{\partial s_{\rm m}} =0
$, in brief,
\begin{equation}
dS=\frac12\,b(\tilde{r}_+)d\tilde{r}_+\,.
\label{diffentropyaction}
\end{equation}
So, $S$ is of the form
\begin{equation}
S=S(\tilde{r}_+)
\,.
\label{totalentropy1}
\end{equation}
We can call $S$
the total
entropy of the semiclassical black hole plus hot thin shell
system and we have found that
it is a function of $\tilde{r}_+$ alone.
We see that Eqs.~\eqref{diffentropyaction}
and \eqref{totalentropy1}
fulfill the condition that the entropy function $S$ must be an exact
differential. Moreover, by integrating Eq.~\eqref{diffentropyaction}
one obtains Eq.~\eqref{totalentropy1}. 
It could
seem that the
total entropy $S$ ought to
depend on the properties of matter in an
arbitrary way, but now we see that this is not so,
it depends on $\tilde{r}_+$, through
the specification of the equation of state 
$b(\tilde{r}_+)$.
The result of Eq.~\eqref{totalentropy1} is
counterintuitive and nontrivial.

One can write Eq.~\eqref{totalentropy1} as $S=S(\tilde{A}_{+})$, where
$\tilde{A}_{+}=4\pi\tilde{r}_+^2$ is the gravitational area, i.e.,
$\tilde{A}_{+}$ is the area corresponding to the gravitational radius
$\tilde{r}_+$.  A particular case would be when $S=\pi\tilde{r}_+^2$,
i.e., $S=\frac14\tilde{A}_{+}$, in which case the system has the
Bekenstein-Hawking entropy formula.

\subsubsection{Matter entropy and black hole entropy}

We have been
interested in making a complete analysis of the total entropy
$S$ of the system and now we
want to determine the entropy of the matter $S_{\rm
m}$ in the thin shell.  
To do it, note that
the total entropy $S$ of the whole system is found by
integration of Eq.~\eqref{diffentropyaction}, specifically, $S=
S_0+
\frac12
\,\int_{r_+}^{\tilde{r}_+} b(r)\,dr$,
where the constant $S_0$ does not depend
on $\tilde{r}_+$, and the lower limit
$r_+$ is chosen for convenience. To find this constant,
we have that when
$r_+=\tilde{r}_+$, there is no shell, and so there is no matter
contribution for the entropy, yielding only black hole entropy, i.e.,
$S_{\rm bh}=\pi r_+^2$. Thus, $S_0=S_{\rm bh}=\pi r_+^2$.
Thus, we can definitely write
\begin{equation}
S_{\rm m}=\frac12\,\int_{r_+}^{\tilde{r}_+}b(r)\,dr\,,
\label{matterentropy}
\end{equation}
which is the matter contribution for the entropy,
and
\begin{equation}
S_{\rm bh}=\pi r_+^2\,,
\label{Sbh}
\end{equation}
which is the black hole Bekenstein-Hawking entropy, indeed, $S_{\rm
bh}=4\pi r_+^2=\frac14 A_+$, with $A_+=4\pi r_+^2$ being the black
hole horizon area, see also \cite{ym,hayward0,can,hayward}.
Then Eq.~\eqref{Swithshellandbh},
which  can be written as
$S=S_{\rm bh}+S_{\rm m}$,
is recovered and properly
interpreted.
Note that in Eqs.~\eqref{matterentropy}
and \eqref{Sbh}
the radius $r_+$ appears, although $S$
in Eq.~\eqref{Swithshellandbh}
is a function of
$\tilde{r}_+$ alone
as stated in Eq.~\eqref{totalentropy1}.
The reason is that $r_+$ is in fact
a constant of integration, which is found
from the ensemble input data.

\section{Black hole surrounded by a hot
self-gravitating thin  shell and a heat
reservoir: Thermodynamic energies and 
temperature analysis
for the system} \label{canonicalenergy}

\subsection{Action $I$ and thermodynamic free energy $F$}

Since $\tilde{k}(R)$ is indeed
$\tilde{k}(R,\tilde{r}_+)$,
one has that Eq.~\eqref{tolman2R},
$\beta=b(\tilde{r}_+)\tilde{k}(R)$,
is an equation for $\tilde{r}_+$ once
the equation of state for $b(\tilde{r}_+)$
has been specified.
Indeed, from Eq.~\eqref{tolman2R} one gets
$\tilde{r}_+=\tilde{r}_+(\beta,R)$, for 
the fixed $\beta$ and $R$.
Moreover, we can now write
the action $I$, Eq.~\eqref{action0}, evaluated at
the critical point as
$I(\beta,R;\tilde{r}_+(\beta,R))
=\beta R\left[1-\tilde{k}(R,\tilde{r}_+(\beta,R))\right]
-S(\tilde{r}_+(\beta,R))$,
so that the action at the critical point is
a function of $\beta$ and $R$, $I=I(\beta,R)$, in brief,
\begin{equation}
I
=\beta R\left[1-\tilde{k}(R)\right]
-S(\tilde{r}_+)
\,,
\label{action1}
\end{equation}
bearing always in mind that
$\tilde{k}(R)=\tilde{k}(R,\tilde{r}_+)$
and $\tilde{r}_+=\tilde{r}_+(\beta,R)$.

Since $I
=\beta F$ and $\beta=\frac1T$, one has that the thermodynamic
free energy $F$ is 
\begin{equation}
F=R\left[1-\tilde{k}(R)\right]
-TS(\tilde{r}_+)
\,,
\label{freeenergyaction2}
\end{equation}
with $F=F(T,R)$, since
$\tilde{r}_+=\tilde{r}_+(T,R)$.
We see that $I$ and $F$ contain the same information,
and one can be traded to the other directly.
We use both, indeed, depending on the context
we use one or the other.

\subsection{Thermodynamic energies}

\subsubsection{Thermodynamic energy $E$ of the system}
From the partition function $Z$
one can find 
the thermal energy of the system,
a thermodynamic quasilocal energy,
using the relation
$E=-\left(\frac{\partial \ln Z}{\partial \beta}\right)_R=
\left(\frac{\partial I}{\partial \beta}\right)_R=
\left(\frac{\partial \beta F}{\partial \beta}\right)_R$.
So from Eq.~\eqref{action1} the thermal energy 
$E$ is
\begin{equation}
E=R\left(1-\tilde{k}(R)\right)\,,
\label{thermalEtotal}
\end{equation}
i.e., $E=R\left(1-\sqrt{1-\frac{\tilde{r}_+}{R}}\right)$,
so that $E=E(R)$, or in full
$E=E(R,\tilde{r}_+)$, with
$\tilde{r}_+=\tilde{r}_+(T,R)$
once the equation of state $b(\tilde{r}_+)$
has been specified.

\subsubsection{Thermodynamic energy $m$ of the matter
and thermodynamic energy $E_{\rm bh}$
of the black hole}

The rest mass energy $m$ of the shell is defined as $ m=4\pi
\sigma_{\rm m} \alpha ^{2}$, where $\sigma_{\rm m}$ is the rest mass
density.  Then, Eq.~\eqref{junctionmass} gives
\begin{equation}
m=\alpha
\left(k(\alpha)-\tilde{k}(\alpha)\right)\,.
\label{junctionmass2}
\end{equation}
The rest mass $m$ is a  function of $\alpha$,
$m=m(\alpha)$, as well as a function of
$\tilde{r}_+$ and $r_+$, so in
full $m=m(r_+,\tilde{r}_+,\alpha)$.
Inverting Eq.~\eqref{junctionmass2} one finds
$\tilde{r}_+=
r_+     +   2m \,k(\alpha)-\frac{m^{2}}{\alpha }
$, or, if one prefers
$\tilde{M}=M+m
\sqrt{1-\frac{2M}{\alpha }}-\frac{m^{2}}{2\alpha }$.
The rest mass energy $m$ is a thermodynamic
energy.

Note that, in addition, 
we can formally
define a black hole thermal energy at $R$
as $E_{\rm bh}(R) =R\left(1- k(R)\right)
=R\left(1-\sqrt{1-\frac{r_+}{R}}\right)$, and
a black hole thermal energy at $\alpha$ can also be
formally defined as 
$E_{\rm bh}(\alpha) =\alpha\left(1- k(\alpha)\right)
=\alpha\left(1-\sqrt{1-\frac{r_+}{\alpha}}\right)$,
i.e.,
\begin{equation}
E_{\rm bh}(R) =R\left(1- k(R)\right)
\,,\quad
E_{\rm bh}(\alpha) =\alpha\left(1- k(\alpha)\right)
\,.
\label{bhenergyformal}
\end{equation}

Then, putting Eq.~\eqref{junctionmass2}
and the  expressions for the black
hole thermal energy
given in Eq.~\eqref{bhenergyformal}
into  Eq.~\eqref{thermalEtotal}
we find
that the energy $E(R)$ can be put in the form
\begin{equation}
E(R)=R-\sqrt{\Bigg[R-E_{\rm bh}(R)-m\Bigg]^{2}
+m^{2}\left(\frac{R}{\alpha }-1\right)
-
-2mR\Bigg[
\frac{E_{\rm bh}(R)}{R}-
\frac{E_{\rm bh}(\alpha)}{\alpha}\Bigg]
}
\,.
\label{energyformal}
\end{equation}
Two features come out of this formula.  One
feature is that it is clear in the
formula that the rest mass energy $m$ of the shell is also related to
a thermal energy at the radius $\alpha$. Defining the thermodynamic
energy at $\alpha$ as $E(\alpha) =\alpha\left(1-
\tilde{k}(\alpha)\right)
=\alpha\left(1-\sqrt{1-\frac{\tilde{r}_+}{\alpha}}\right)$, one sees
from Eqs.~\eqref{junctionmass2}
and \eqref{bhenergyformal} that $m$ can be written as $m\equiv
E(\alpha)- {E_{\rm bh}(\alpha)}$, a fact that can be also found by
substituting $\alpha$ for $R$ in the above expression for $E(R)$.  The
other feature is that the total energy $E$ at $R$ is not a sum of the
black hole energy $E_{\rm bh}(R)$, plus the rest mass energy $m\equiv
E(\alpha)- {E_{\rm bh}(\alpha})$, it is a nonlinear function of both,
in fact, the energies $E(R)$, $m\equiv E(\alpha)- {E_{\rm
bh}(\alpha)}$, and
${E_{\rm
bh}(R)}$
are related in a rather complicated way.

\subsection{Temperature analysis}

\subsubsection{Reduced temperature equation of state for the
thermodynamic system}

To start the temperature analysis we reiterate that from the condition that
the main contribution to the path integral, and so to the partition
function $Z$, comes indeed from the classical action, as it is
required in the semiclassical approximation, i.e., $dI=0$, and from
the ascertainment that $S$ is an exact differential, we deduced
Eq.~\eqref{tolman2R}, i.e., $\beta=b(\tilde{r}_+)\tilde{k}(R)$ with
$b(\tilde{r}_+)$ being a function of $\tilde{r}_+$ alone.  Since
$\beta=\frac1T$, see Eq.~\eqref{inverseT}, we can put
Eq.~\eqref{tolman2R} in terms of the temperature, which is useful,
\begin{equation}
T=\frac{T_{0}(\tilde{r}_+)}{\tilde{k}(R)}\,,
\quad\quad
T_0(\tilde{r}_+)\equiv\frac{1}{b(\tilde{r}_+)}
\label{binverseT0}\,,
\end{equation}
with $T_{0}(\tilde{r}_+)$ being a function of $\tilde{r}_+$ alone, but
otherwise free. We have called $b(\tilde{r}_+)$ the reduced inverse
temperature equation of state and, similarly, we call
$T_{0}(\tilde{r}_+)$ the reduced temperature equation of state.  
Of course, Eqs.~\eqref{tolman2R} and \eqref{binverseT0}
are equivalent and we will
choose to use one or the other in accord
with convenience.

This result, Eq.~\eqref{binverseT0}, or Eq.~\eqref{tolman2R}, is
essential for the analysis of the temperature.
As we have noted in relation to 
Eq.~\eqref{tolman2R}, we can now emphasize that since
$T$ and $R$ are fixed, because these
are the fixed quantities that
characterize the ensemble, and
$T_0$ is a function of $\tilde{r}_+$ alone,
$T_0=T_0(\tilde{r}_+)$, one
has that Eq.~\eqref{binverseT0} is an
equation for the possible gravitational radii $\tilde{r}_+$ of the
space,
$\tilde{r}_+=\tilde{r}_+(T,R)$.
 For
instance, it is conceivable that for the same boundary data
$T$ and $R$, 
there is
the possibility of existing one solution $\tilde{r}_{+1}(T,R)$
with a small $\tilde{r}_+$
and other solution  $\tilde{r}_{+2}(T,R)$
with a large $\tilde{r}_+$. It is also conceivable
the existence of three or more solutions for $\tilde{r}_+(T,R)$.

\subsubsection{Temperature relation between boundary and black hole}

In the temperature analysis one should
note that there are two important
requirements that come out when
setting the path integral to find the
reduced action given in
Eq.~\eqref{action0}.
The first requirement is that the temperature, or
the inverse temperature, is fixed at the boundary $R$.
This implies that, since
we are dealing with a Euclideanized time with the
$S^1$ at the
boundary with radius $R$ having length $\beta$, one can write
$\beta=P\tilde{k}(R)$ for an arbitrary fixed period $P$.
The second requirement,
is that in order that the Euclidean
two-manifold does not have
conical singularities, one must impose
that
$\left[ {k}(r) {k}^\prime(r)
\dfrac{ \tilde{k}(\alpha)}
{{k}(\alpha)} \right]_{r_+}
=\dfrac{2\pi}{P}$.
These two equations
coming out of the two requirements
together, plus $\beta=\frac1T$, yield 
\begin{equation}
4\pi r_+k(\alpha)=
\frac{1}{T}\frac{\tilde{k}(\alpha)}
{\tilde{k}(R)}\,,
\label{rt}
\end{equation}
or, $4\pi r_+\sqrt{1-\frac{r_+}{\alpha }}=
\frac{1}{T}\frac{\sqrt{1-\frac{ \tilde{r}_+}{\alpha }}}{
\sqrt{1-\frac{\tilde{r}_+}{R}}}$.  Since the temperature $T$ and the
radius $R$ are fixed, it implies that the choice of the radii
$\tilde{r}_+$ and $r_+$ is not arbitrary, there is a relation imposed
on them together with $\alpha$.  From Eq.~\eqref{rt} we find $r_+$,
namely, $r_+=r_+(\tilde{r} _+,T,\alpha)$, but as we have seen
$\tilde{r}_+=\tilde{r}_+(R,T)$, see Eq.~\eqref{binverseT0} or
equivalently Eq.~\eqref{tolman2R}, so that $r_+=r_+(\tilde{r}
_+,T,\alpha)=r_+(\tilde{r}_+(R,T),T,\alpha)=r_+(R,T,\alpha)$, i.e.,
the black hole horizon radius $r_+$ is found from the boundary data
plus $\alpha$, once the equation of state
for the reduced temperature 
$T_0(\tilde{r}_+)$ is specified.

Incidentally, note from Eq.~\eqref{rt} that the limit of no black hole
inside the shell, i.e., $r_+=0$, cannot be taken here, since the left
hand side of Eq.~\eqref{rt} goes to zero, i.e., the Hawking
temperature of the black hole inside, which is equal to $\frac1{4\pi
r_+}$, diverges.  This is a sign that the disappearance of a black
hole leaves a defect in the space rather than pure flat space, at
least at this order of approximation.
In turn, this means that the treatment of a shell with
nothing, i.e., flat space, inside in
a heat reservoir has to be done
independently. 
We will not do it here, but
this solution will be mentioned, and
will make its first appearance when we give an exact solution
for the ensemble below.

\subsubsection{Tolman relations for the temperature}

Some other properties of the temperature are worth
mentioning which reflect further the temperature
stratification in the system.

For the region outside of the thin shell, the Tolman relation states
that for some radius $r$, $T(r)\tilde{k}(r)=T(R)\tilde{k}(R)$, so that
\begin{equation}
T(r)=T\frac{\tilde{k }(R)}{\tilde{k}(r)}\,,\quad\quad 
   \alpha\leq r \leq R\,,
\label{tolmanoutr}
\end{equation}
where $T$ is the fixed temperature at $R$, $T\equiv T(R)$.
From Eq.~\eqref{binverseT0}, we can put 
Eq.~\eqref{tolmanoutr} as
$T(r)=\frac{T_{0}(\tilde{r}_+)}{\tilde{k}(r)}$, and since
at infinity $\tilde{k}(r)$ is equal to one,
we now see
that one can think of the reduced temperature $T_{0}(\tilde{r}_+)$ of
Eq.~\eqref{binverseT0} as having the meaning of a temperature at
infinity, in the sense that if some imagined radiation escaped from
the system to infinity that very radiation would have temperature
$T_{0}(\tilde{r}_+)$ at infinity, and so in this analogy it has an
equivalent meaning to the Hawking temperature for a black hole which
is defined as its temperature at infinity.
Now, Eq.~\eqref{tolmanoutr} can be evaluated at the shell,
i.e., at $\alpha$, to give
\begin{equation}
T(\alpha)=T\frac{\tilde{k }(R)}{\tilde{k}(\alpha)}
\,,\quad\quad 
   \alpha= r \,,   
\label{tolmanoutalpha}
\end{equation}
i.e., $T(\alpha)=T\frac{ \sqrt{ 1-\frac{\tilde{r}_+}{R} }} {\sqrt{
1-\frac{\tilde{r}_+}{\alpha} }}$. This is the expression for the
temperature at $\alpha$ found from the outside region.  Since
$\alpha\leq R$ one has from Eq.~\eqref{tolmanoutalpha} that
$T(\alpha)\geq T$, i.e., the temperature of the shell is greater than
the temperature of the reservoir.

For the region inside of the thin shell,
we first use the relation
\begin{equation}
T(\alpha)=\frac{1}{4\pi r_+k(\alpha )}
\,,\quad\quad 
   r=\alpha \,, 
\label{Talpha1}
\end{equation}
i.e.,
$T(\alpha)=\frac{1}{4\pi r_+\sqrt{1-\frac{\tilde{r}_+}{\alpha}}}$,
which is found directly from 
Eq.~\eqref{rt} together with Eq.~\eqref{tolmanoutalpha}.
Then using again the Tolman relation, i.e.,
$T(r){k}(r)=T(\alpha){k}(\alpha)$,
one finds 
\begin{equation}
T(r)=\frac{1}{4\pi r_+k(r)}\,,\quad\quad r_+\leq r\leq\alpha
\,, \label{localTfromi}
\end{equation}
which gives the temperature stratification in the inside region.

\subsubsection{A further important relation}

Using Eq.~\eqref{binverseT0}, or Eq.~\eqref{tolman2R}, specifically,
$\frac1T=b(\tilde{r}_+)\tilde{k}(R)$,
together with
Eq.~\eqref{rt}, i.e., $4\pi r_+k(\alpha)=
\frac{1}{T}\frac{\tilde{k}(\alpha)} {\tilde{k}(R)}$, one has
\begin{equation} 4\pi
r_+k(\alpha)= b(\tilde{r}_+)\tilde{k}(\alpha)\,,
\label{rtdifferent}
\end{equation}
i.e., $4\pi r_+\sqrt{1-\frac{r_+}{\alpha }}
=b(\tilde{r}_+)\sqrt{1-\frac{\tilde{r}_+}{\alpha}}$.
From Eq.~\eqref{rtdifferent} one finds
$b(\tilde{r}_+)\geq 4\pi r_+$.

Now, Eq.~\eqref{rtdifferent} is important in several settings, in
particular, one can make a preliminary analysis of the possible high
temperature limits.  It
is advisable to divide
the analysis into two main cases, the
first one being finite heat reservoir temperature $T$ and the second
one being diverging to infinite heat reservoir temperature $T$.
In
the first case, if the heat reservoir temperature $T$ is finite and
the shell temperature $T(\alpha)$ is very high, we have from
Eqs.~\eqref{Talpha1}
and Eq.~\eqref{rtdifferent},
that $T(\alpha)$ very high means that there are two subcases: either
$\tilde{r}_+\to\alpha$ and $r_+\to\alpha$, which remarkably implies
$b(\alpha)\geq 4\pi \alpha$ for sure, or $\tilde{r}_+\to\alpha$ and
$r_+\to0$.
In
the second case, if the heat reservoir temperature $T$ is very high
then the shell temperature $T(\alpha)$ is also very high, there are
also two subcases, see  Eq.~\eqref{binverseT0}, or
Eq.~\eqref{tolman2R},
and Eq.~\eqref{rtdifferent}: either $\tilde{r}_+\to\alpha\to R$ and
$r_+\to0$ or $b(\tilde{r}_+)\to4\pi\tilde{r}_+\to0$ and $r_+\to0$.
A complete study of the high temperature limit
with the presentation of analytical solutions is made below.

\section{First law of thermodynamics for black hole and hot
slef-gravitating thin shell
in a heat reservoir system,
consistency of the whole scheme, and thermodynamic stability}
\label{firstlaw}

\subsection{First law of thermodynamics}
\subsubsection{First law of thermodynamics for the system}

The first law of thermodynamics in a shortened form, i.e.,
$TdS=dE$, has indeed emerged from the minimum of the action principle
which applies to systems with constant temperature and constant area.
Indeed, with the help of the definition of $E$ given in
Eq.~\eqref{thermalEtotal}, the reduced action $I$ can now be written as
$I=\beta E-S$, and so at constant $T$ and constant $R$, the minimum of
the action $dI=0$ occurs for $\beta dE-dS=0$. Since $\beta=\frac1T$,
one finds the first law of thermodynamics in the shortened form
$TdS=dE$, where $dE=\left(\frac{\partial E}{\partial {\tilde
r}_+}\right)_R\, d{\tilde r}_+$ as $R$ is kept constant.

But once the minimum is found and applied back to the action one has
that the action is now a function of $\beta$ and $R$ alone, i.e.,
$I=I(\beta,R)$, see Eq.~\eqref{action1}, namely,
$I=I(\beta,R,\tilde{r}_+)$ with $\tilde{r}_+=\tilde{r}_+(\beta,R)$.
Thus, with $\beta=\frac1T$, the thermodynamic free energy $F$
also has a functional dependence of the form
$F=F(T,R)$, see Eq.~\eqref{freeenergyaction2},
or more conveniently
$F=F(T,A)$, with $A=4\pi R^2$.
At this stage, one is outside the regime
of ensembles and statistical mechanics,
and is in the domain of thermodynamics.
So, one can compare systems of the same type
that change infinitesimally in temperature, $dT$, and
infinitesimally
in area, $dA$. Since $F$ characterizes the system,
we look for  changes $dF$ related to changes
$dT$ and $dA$, i.e.,
$dF=\left(\frac{\partial F}{\partial T}\right)_AdT+
\left(\frac{\partial F}{\partial A}\right)_TdA$.
Now, from thermodynamics one has 
$\left(\frac{\partial F}{\partial T}\right)_A=-S$
and $\left(\frac{\partial F}{\partial A}\right)_T=-p$,
where $p$ is the thermodynamic
tangential pressure at the boundary with area $A$.
So, $dF=-SdT-pdA$.
On the other hand, from
Eq.~\eqref{freeenergyaction2}
together with Eq.~\eqref{thermalEtotal}, one
has $F=E-TS$, so that $dF=dE-TdS-SdT$.
These two latter expression for $dF$
together give then
\begin{equation}
TdS=dE+pdA,
\label{1w}
\end{equation}
which is the usual first law
of thermodynamics for the system.
The thermodynamic
tangential pressure $p$ at the boundary
can defined in several ways, namely,
$p=-\left(\frac{\partial F}{\partial A}\right)_T$,
$p=T\left(\frac{\partial S}{\partial A}\right)_T$,
or
$p=\left(\frac{\partial E}{\partial A}\right)_S$.
Let us use the latter relation,
and since constant $S$ is equivalent to constant $\tilde r_+$,
see Eq.~\eqref{totalentropy1},
one has
$p=-\left(\frac{\partial E}{\partial A}\right)_{\tilde r_+}$,
and so from Eq.~\eqref{thermalEtotal} one gets
\begin{equation}
p(R)=
\frac1{16\pi R}
\frac{\Big(1-\tilde{k}(R)
\Big)^{2}}{\tilde{k}(R)}\,.
\label{pressuresystem}
\end{equation}
This thermodynamic gravitational  pressure applies
tangentially at all points of the
reservoir with area $A$.

One can now verify that the first law of thermodynamics for the system
given in Eq.~\eqref{1w} holds true.  One inserts the differential form
of $S$ from Eq.~\eqref{diffentropyaction}, takes the differential of
the expression for $E$ given in Eq.~\eqref{thermalEtotal}, inserts the
pressure term $p$ of Eq.~\eqref{pressuresystem}, and takes the
differential of the expression for $A$ given in Eq.~\eqref{areacavity},
to find that Eq.~\eqref{1w} is an identity.

\subsubsection{First law of thermodynamics for the hot matter
in the thin shell
and first law of thermodynamics for the black hole}

We are now interested in finding whether we can apply the first law of
thermodynamics to the hot shell itself at $\alpha$ and also apply the
first law to the black hole.  As the energies $E(R)$, $E(\alpha)\equiv
m$, and $E_{\rm bh}(R)$ and $E_{\rm bh}(\alpha)$, are related in a
rather complicated nonlinear way, and are calculated at different
radii, namely at $R$ and at $\alpha$, see Eq.~\eqref{thermalEtotal},
the answer is not obvious.

Applying formally the first law of thermodynamics to the shell at
$\alpha$ one has
\begin{equation}
T_{\rm m}\,dS_{\rm m}=
dm+
p_{\rm m}\,dA_{\rm m}\,,
\label{sm}
\end{equation}
where $T_{\rm m}$ is the temperature of the matter in the shell, with
$T_{\rm m}\equiv T(\alpha)$, see also Eqs.~\eqref{tolmanoutalpha} and
\eqref{Talpha1}, $S_{\rm m}$ is the entropy of the matter in the
shell, with $S_{\rm m}\equiv S(\alpha)$, see also
Eq.~\eqref{matterentropy}, $m$ is the rest mass energy density of the
matter in the shell, given by $m=4\pi \sigma_{\rm m} \alpha^2$, see
also Eq.~\eqref{junctionmass2}, $p_{\rm m}$ is the tangential pressure
of the matter in the shell, with $p_{\rm m}\equiv p_{\rm m}(\alpha)$,
and is given by Eq.~\eqref{pressshell1}, and $A_{\rm m}$ is the area
of the shell, with $A_{\rm m}\equiv A(\alpha)=4\pi \alpha^2$.
Thus, from Eq.~\eqref{junctionmass2} we have $m=\alpha
\left(k(\alpha)-\tilde{k}(\alpha)\right) $, where
$k(\alpha)=\sqrt{1-\frac{r_+}{\alpha}}$ and
$\tilde{k}(\alpha)=\sqrt{1-\frac{\tilde{r}_+}{\alpha}}$.  The
tangential pressure at the shell, at $\alpha$, is given by $ 8\pi
p_{\rm m}=\frac{\tilde{k}(\alpha)-k(\alpha)}{\alpha }+ \left(
\tilde{k}^{\prime}(\alpha) -{k}^{\prime}(\alpha)\right) \label{pa3} $,
which follows from the junction conditions, see
Eq.~\eqref{pressshell1}.  A direct calculation shows that $dm+p_{\rm
m}\,dA_{\rm m}= \frac{1}{2\tilde{k}(\alpha)} d\tilde{r}_+
-\frac{1}{2k(\alpha)}dr_+$.  Now, from Eq.~\eqref{tolmanoutalpha} and
Eq.~\eqref{tolman2R} one has $T_{\rm m}=
T(\alpha)=\frac{T_{0}(\tilde{r}_+)}{\tilde{k}(\alpha )}$, and from
Eq.~\eqref{Talpha1} one has $T_{\rm m}= T(\alpha)=\frac{1}{4\pi
r_+k(\alpha)}$, so that $\frac1{T_{\rm m}}\Big( dm+ p_{\rm m}\,dA_{\rm
m}\Big) = \dfrac{d\tilde{r}_+}{2T_{0}} -2\pi r_+dr_+ $.  But the first
term of this equality, is the differential of the total entropy $dS$
as given by $dS=\frac12 b(\tilde{r}_+)d\tilde{r}_+$, see
Eq.~\eqref{diffentropyaction}, and the second term of this equality is
the differential of the Bekenstein-Hawking entropy for the black hole,
$dS_{\rm bh}=2\pi r_+ dr_+$, see Eq.~\eqref{Sbh}.  Thus, we deduce
that $dS_{\rm m}=dS-dS_{\rm bh} \label{additivityS} $, a result
showing that the differential of the entropies are additive,
recovering Eq.~\eqref{Swithshellandbh} in differential form.  Thus,
indeed, the first law for the matter at the shell, at $\alpha$, as
given in Eq.~\eqref{sm} holds true.

It is also worth to find whether we can apply the first law
of thermodynamics
to
the black hole itself considering that there is an effective cavity
at $\alpha$. In this case, the first law for the black hole would be
\begin{equation}
T(\alpha)dS_{\rm bh}=
dE_{\rm bh}(\alpha)+p_{\rm m}(\alpha)dA(\alpha)\,,
\label{firstlawbh1}
\end{equation}
where the functional dependence in
$\alpha$ is explicitly written to indicate
that the quantities in the stated first law
of thermodynamics, Eq.~\eqref{firstlawbh1}, are evaluated at
the cavity at radius $\alpha$.
The black hole energy at $\alpha$
has been defined in Eq.~\eqref{bhenergyformal}
as $E_{\rm bh}(\alpha)
=\alpha\left(1- k(\alpha)\right)
=\alpha\left(1-\sqrt{1-\frac{r_+}{\alpha}}\right)$.
Using this expression in Eq.~\eqref{firstlawbh1}
one indeed finds
that $ dS_{\rm bh}=2\pi r_+dr_+$, making the analysis of
the previous paragraph that yielded $ dS_{\rm m}=dS-dS_{\rm bh}$
consistent. Thus, Eq.~\eqref{firstlawbh1} is a form of the first law 
of thermodynamics
for the part of the system inside the hot
shell where a black hole is
situated.

We have thus three versions of the first law of thermodynamics.  One
is for the whole system, see Eqs.~\eqref{1w} plus the result
given in
\eqref{pressuresystem}, another for the shell itself, see
Eq.~\eqref{sm}, and yet a third for the part of the system inside the
shell where a black hole is situated, see Eq.~\eqref{firstlawbh1}.
This is remarkable.  Moreover, this happens in spite of the fact that
the energies are not additive.

\subsection{Consistency of the whole scheme}

Now that we are in
possession of all the necessary data we can show 
that the whole scheme is consistent.

\noindent
(i) Set a canonical ensemble for the system, i.e., give a spherical
reservoir at temperature $T$ and radius $R$.

\noindent
(ii) Choose a specific equation of state for the reduced inverse
temperature $b(\tilde{r}_+)$, or, which is the same thing, choose a
specific equation of state for the reduced temperature
$T_0(\tilde{r}_+)$.

\noindent
(iii) Then from Eq.~\eqref{tolman2R}, or Eq.~\eqref{binverseT0}, find the
possible gravitational radii $\tilde{r}_+=\tilde{r}_+(T,R)$ compatible
with the boundary values $T$ and $R$.

\noindent
(iv) On the other hand, the gravitational radius $\tilde{r}_+$ has the
functional dependence $\tilde{r}_+=f(r_+,\alpha,\sigma_{\rm m})$, since from
Eq.~\eqref{junctionmass2} one has $\tilde{r}_+= r_+ + 2m
\,k(\alpha)-\frac{m^{2}}{\alpha } $.

\noindent
(v) Now, we have seen that from Eq.~\eqref{rt} one finds
$r_+=r_+(T,R,\alpha)$, i.e., the black hole horizon radius $r_+$ is
found from the boundary data plus $\alpha$.

\noindent
(vi) Thus, since $\tilde{r}_+=f(r_+,\alpha,\sigma_{\rm m})$ one has
$\tilde{r}_+=\bar f(T,R,\alpha,\sigma_{\rm m})$, for some function
$\bar f$.

\noindent
(vii) But, since $\tilde{r}_+=\tilde{r}_+(T,R)$ and $\tilde{r}_+=\bar
f(T,R,\alpha,\sigma_{\rm m})$, one has $\tilde{r}_+(T,R)=\bar
f(T,R,\alpha,\sigma_{\rm m})$.

\noindent
(viii) So, for given boundary data $T$ and $R$,
one finds $\sigma_{\rm m}=g_{\rm m}(\alpha)$,
for some function $g_{\rm m}$, i.e.,
one finds
an equation giving the energy density
$\sigma_{\rm m}$ of the matter in the shell
as a function of the radius $\alpha$ of the shell.

\noindent
(ix) One can now think of two possibilities to proceed
to close the system.

\noindent
(ixa)
A possibility which is quite natural is to assume that there is an
equation of state for the matter on the shell of the form
 $\sigma_{\rm m}=\sigma_{\rm
m}(s_{\rm m},\alpha)$, or if one prefers, of the form
$s_{\rm
m}=s_{\rm m}(\sigma_{\rm m},\alpha)$.
Then, the equation of state for
$p_{\rm m}$ is automatically given because $p_{\rm m}=
-\left(\frac{\partial m}{\partial A_{\rm m}}\right)_{S_{\rm m}}
=
-\left(\frac{\partial
(\sigma_{\rm m}\alpha^2)}{\partial \alpha^2}\right)_{S_{\rm m}}$,
see Eq.~\eqref{sm}.  So, we can
write $p_{\rm m}=\pi(\sigma_{\rm m},\alpha)$, for some function $\pi$.
But
$p_{\rm m}=p_{\rm m}(\tilde{r}_+,r_+,\sigma_{\rm m},\alpha)$
from the junction condition, see
Eq.~\eqref{pressshell1}, and so $p_{\rm m}=\bar
p_{\rm m}(R,T,\sigma_{\rm m},\alpha)$, for
some function $\bar p_{\rm m}$, from the discussion above.
So, for fixed $T$
and $R$, $p_{\rm m}=\bar p_{\rm m}(\sigma_{\rm m},\alpha)$.  Then,
$\pi(\sigma_{\rm m},\alpha)=\bar p(\sigma_{\rm m},\alpha)$,
which yields $\sigma_{\rm m}=\bar
\sigma_{\rm m}(\alpha)$, for some function $\bar \sigma_{\rm m}$.  But
$\sigma_{\rm m}=g_{\rm m}(\alpha)$ as was found independently
above.  This means
that the following equality holds, $g_{\rm m}(\alpha)=\bar
\sigma_{\rm m}(\alpha)$, which is an equation to obtain $\alpha$.
Once one has
$\alpha$, then one obtains the other
important quantities, and the solution is closed
with values for
$\tilde{r}_+$, $r_+$, $\sigma_{\rm m}$,
$p_{\rm m}$,
and $s_{\rm m}$.

\noindent
(ixb)
Another possibility which is more straightforward is to give the
radius $\alpha$ of the shell, i.e., fix $\alpha$ from the start. Then
everything can be
calculated from the input data
directly, namely,
one finds $\tilde{r}_+$, $r_+$, $\sigma_{\rm m}$,
$p_{\rm m}$, and $s_{\rm m}$.  In this case the shell radius $\alpha$
is a free parameter that upon specification solves the problem.  Thus,
to give $\alpha$ works in practical terms like giving an equation
of state for the matter. In any case,
we will see that $\alpha$ can be constrained
to lie with some range of values of $\tilde{r}_+$, say, when a
specific solution is known.

\subsection{Thermodynamic stability of the system black hole
and hot thin shell
inside a heat reservoir}
\label{stability}

Thermodynamic stability of a
thermodynamic system is always an important issue.  In
this case it hinges on the heat capacity.  The heat capacity at
constant reservoir area $A$ is defined by $C_A= \left(\frac{\partial
E}{\partial T}\right)_A$, and thus thermodynamic stability
exists when
\begin{equation}
C_A\geq0,
\label{heatCstabil}
\end{equation}
i.e, to be stable it means that if the system receives energy its
temperature must increase.  Now, constant $A$ is also constant $R$, so
one has $C_A= \left(\frac{\partial E}{\partial T}\right)_A=
\left(\frac{\partial E}{\partial T}\right)_R$, and so from
Eq.~\eqref{thermalEtotal} $C_A$ is $C_A= \left(\frac{\partial
E}{\partial T}\right)_R= \frac{1}{2\tilde{k}} \left( \frac{\partial
\tilde{r}_+}{\partial T}\right)$. Then, from Eq.~\eqref{binverseT0},
i.e., $T=
\frac{T_{0}(\tilde{r}_+)}{\sqrt{1-\frac{\tilde{r}_+}{R}}}$, one
finds $C_A=\frac{R \left(1-\frac{ \tilde{r}_+}{R}\right)} {
T_0+2T_0^{\prime}R \left(1-\frac{\tilde{r}_+}{R}\right)}$, where a
prime means derivative to its own argument.  It is more useful to use
the inverse temperature $b(\tilde{r}_+)$ defined in
Eq.~\eqref{binverseT0}, i.e., $b(\tilde{r}_+)=
\frac1{T_{0}(\tilde{r}_+)}$. Then one has for the  heat capacity
\begin{equation}
C_A=\frac{b^2(\tilde{r}_+) R\,
{\tilde{k}}^2(R)
}
{b(\tilde{r}_+)-2b^{\prime}(\tilde{r}_+)\, R\,
      {\tilde{k}}^2(R)},
\label{heatC}
\end{equation}
or, putting the redshift factors
in explicit form, one
has
$
C_A=\frac{b^2(\tilde{r}_+) R \left(1-\frac{ \tilde{r}_+}{R}\right)}
{b(\tilde{r}_+)-2b^{\prime}(\tilde{r}_+) R
\left(1-\frac{\tilde{r}_+}{R}\right)}
$.
Clearly, the thermodynamic condition for stability,
Eq.~\eqref{heatCstabil}, is
essentially given by $b(\tilde{r}_+)-2b^{\prime}(\tilde{r}_+)R
\left(1-\frac{\tilde{r}_+}{R} \right)\geq0$,
which can be put in the form
\begin{equation}
\tilde{r}_+\leq
R\leq \tilde{r}_++
\frac{b(\tilde{r}_+)}{2b^{\prime}(\tilde{r}_+)}.
\label{Rstablecond}
\end{equation}
Interesting to note
that for an equation of state $b(\tilde{r}_+)$ that obeys
the condition 
${b^{\prime}}^2\geq b b^{\prime\prime}$, one can find from
Eq.~\eqref{Rstablecond} that $R$ also obeys a secondary inequality
$R\leq \alpha+\frac{b(\alpha)}{2b^{\prime}(\alpha)}$.

The outcome of the stability analysis for this system is that given a
specific equation of state $b(\tilde{r}_+)$ one finds from
Eq.~\eqref{tolman2R} the possible $\tilde{r}_+$ compatible with the
boundary data $\beta$, or $T$,
and $R$. There can exist many, and then one
verifies which of the $\tilde{r}_+$ found are stable through
Eqs.~\eqref{heatC} and \eqref{Rstablecond}.  As an example, assume that
the equation of state for the reduced inverse temperature $b(\tilde{r}_+)$
has a power law form, i.e., $b(\tilde{r}_+)={\gamma \ell}\left(
\frac{\tilde{r}_+}{\ell} \right)^{\hskip -0.1cm\delta}$, for some
$\gamma$, $\ell$, and $\delta$, constant quantities that are intrinsic
to the system properties, with 
$\gamma$ and $\delta$ without units, and $\ell$ with units
of length,
obeying
$0\leq\gamma<\infty$, $0\leq\ell<\infty$, and
$0<\delta<\infty$.  Depending on $\gamma$, $\ell$,
and $\delta$ the power-law equation
of state can have one, two, three, or more solutions
for $\frac{\tilde{r}_+}{\ell}$
for given reservoir  $\beta$, or $T$, 
and $R$.  From Eq.~\eqref{heatC}
one finds that the heat capacity in this case
is $C_A= \gamma
\ell^2\left(\frac{\tilde{r}_+}{\ell}\right)^{\delta +1}
\frac{\left(1-\frac{
\tilde{r}_+}{R}\right)}{\Big[\frac{\tilde{r}_+}{R }\left(1+2\delta
\right)-2\delta\Big]}$, and Eq.~\eqref{Rstablecond} gives
$\tilde{r}_+\leq
R\leq\frac{1+2\delta}{2\delta}\, \tilde{r}_+$.  It comes out from
this equation that for very small $\delta$, $\delta\to0$, one gets
$1\leq \frac{R}{\tilde{r}_+}<\infty$, i.e., any $R$ gives a
thermodynamically stable solution, whereas in the other limit of very
high $\delta$, $\delta\to\infty$, one gets
$1\leq \frac{R}{\tilde{r}_+}\leq 1$,
i.e., 
$\frac{R}{\tilde{r}_+}= 1$,
meaning that the only thermodynamically
stable situation is the case in which the reservoir radius $R$ is at
the gravitational radius $\tilde{r}_+$ of the system, which since
$\tilde{r}_+\leq\alpha$ it has also to be the radius $\alpha$ of the
shell. Values of $\delta$ intermediate between these two limiting
values give correspondingly intermediate $\frac{R}{\tilde{r}_+}$
values.

It is worth noting that the thermodynamic condition for stability,
essentially given by $b(\tilde{r}_+)-2b^{\prime}(\tilde{r}_+)R
\left(1-\frac{\tilde{r}_+}{R} \right)\geq0$, see
Eqs.~\eqref{heatCstabil}-\eqref{Rstablecond}, is determined in terms
of quantities that are given at the boundary, i.e., $\beta$, or $T$,
and $R$, or if one prefers, in terms of $R$ and of the
equation of state $b\left(\tilde{r}_+\right)$, since
$\beta={b(\tilde{r}_+)k(R)}$, see Eq.~\eqref{tolman2R}.
As a
result, even if the black hole inside the shell is small, i.e., $r_+$
is small, the system can nevertheless be stable.  This is remarkable
and there is no analogue in York's case, the case when a pure black
hole, i.e., a black hole without a shell, is situated inside a
cavity. In that pure black hole case, only the larger black hole, the
black hole with large mass, is stable.  To understand heuristically
this new result of stability of a small black hole, suppose the system
is thermodynamically stable.  Thus, under some quite general
condition, this means that the heat reservoir radius obeys $ R\leq
\tilde{r}_++ \frac{b(\tilde{r}_+)}{2b^{\prime}(\tilde{r}_+)}$,
see Eq.~\eqref{Rstablecond}, i.e., $\tilde{r}_+$ is some fraction of
$R$ in any concrete instance.
It then happens that a small decrease in the small black hole
energy and so in its mass, or what is the same, in $r_+$, would be
accompanied by an increase in the black hole temperature, which can be
thought of as an increase in the radiation emitted.  However, due to
this increase of black hole temperature and radiation, the system as a
whole increases its temperature, and so its energy also increases
since the system is supposed stable.  The increase in energy of the
whole system then compensates for the decrease in the inner black hole
mass, and so the consequence is that the black hole gets back more
energy than the quantity it has delivered initially, and finally the
black hole cools slightly, returns to equilibrium, keeping the
system stable. Moreover, this phenomenon happens for any of the
possible black hole solutions $r_+$ for which the
system is thermodynamically stable, 
i.e., the solutions of the equation $r_+=r_+(R,T,\alpha)$, see
Eq.~\eqref{rt} and the discussion after it, that
might arise in a thermodynamically stable system. Of course, if the
temperature at $\alpha$ were fixed then one would fall into York's
case, the shell would act as a heat reservoir, and of the now two
possible solutions, $r_{+1}$ would be unstable and $r_{+2}$ would be
stable, but the point here is that what is fixed is the temperature of
the canonical ensemble at $R$, the temperature of
the hot shell at
$\alpha$ is not strictly
fixed,
and the same happens to the temperatures in the region
between the shell and the heat reservoir,
and this fact allows for the stability of the
black hole solutions inside the shell, be they small or large.

\section{Exact thermodynamic solutions for a specific equation of
state: Hot self-gravitating thin shell, pure black hole, and hot flat
space phases of the canonical ensemble}
\label{specificeos}

\subsection{Specific equation of state, action,
entropy, and heat capacity}
\label{a}

To make progress we need to specify an explicit equation of state for
the reduced inverse temperature 
$b(\tilde{r}_+)$. One quite interesting
equation of state would be the
one
where
$b(\tilde{r}_+)={\gamma \ell}\left( \frac{\tilde{r}_+}{\ell}
\right)^{\hskip -0.1cm\delta}$, with $\gamma$, $\ell$,
and $\delta$ as free
parameters. This equation of state is still too general for our
purposes of solving  the
microscopic gravitational system
composed of reservoir, hot shell, and black hole
in a relatively simple but meaningful
and interesting way.
To proceed, note that
a system with black hole
quantum thermal properties, but
not necessarily a black hole, has its inverse temperature
$b(\tilde{r}_+)$ proportional to the inverse Hawking temperature,
i.e., $b(\tilde{r}_+)=\gamma \tilde{r}_+$, in which case
the exponent $\delta$ is unity, $\delta=1$,
and moreover,
if the system's reduced
temperature has exactly the Hawking expression,
then the parameter $\gamma$ has a precise value, namely,
$\gamma=4\pi$. A system with $\delta=1$ and $\gamma=4\pi$ yields
the simplest nontrivial equation of state for
the reduced temperature.  So,
the equation of state $b(\tilde{r}_+)$ we choose for the 
system is
\begin{equation}
b(\tilde{r}_+)=4\pi \tilde{r}_+.
\label{bhawktilder}
\end{equation}
In brief, we give as equation of state for the
reduced inverse temperature of
the system precisely the Hawking form. Note, that although the
equation of state of Eq.~\eqref{bhawktilder} has the Hawking inverse
temperature form, $\tilde{r}_+$ is the gravitational radius of the
shell, and so here, in general, Eq.~\eqref{bhawktilder} is not an
equation for a black hole.

The action $I$ of q.~\eqref{action0},
or of 
Eq.~\eqref{action1}, for the equation of state
given in Eq.~\eqref{bhawktilder} is
\begin{equation}
I=\beta R\left[1-\tilde{k}(R)\right]
-\pi \tilde{r}_+^2,
\label{actionspecific}
\end{equation}
with $\tilde{r}_+=\tilde{r}_+(T,R)$.
The free energy
given in Eq.~\eqref
{freeenergyaction2}
is therefore $F= R\left[1-\tilde{k}(R)\right]
-T\pi \tilde{r}_+^2$.
The entropy $S$
of the system given in Eq.~\eqref{Swithshellandbh}
for the equation of state
Eq.~\eqref{bhawktilder}
turns into $S=\pi \tilde{r}_+^2$, i.e.,
\begin{equation}
S=\frac14 \tilde{A}_{+}\,,
\label{entropyspecific}
\end{equation}
where $\tilde{A}_{+}= 4\pi \tilde{r}_+^2$, and is
thus the Bekenstein-Hawking entropy for
the gravitational area of the shell, which again in general is not a
black hole.
The entropy of the interior black hole
is still $S_{\rm bh}=\pi r_+^2=\frac14 A_+$.
The matter entropy $S_{\rm m}$ of Eq.~\eqref{matterentropy} for the
equation of state given in Eq.~\eqref{bhawktilder} is
$S_{\rm m}=\pi\left( \tilde{r}_+^2-r_+^2\right)$, i.e.,
\begin{equation}
S_{\rm m}=\frac14\left( \tilde{A}_{+}-A_+\right)\,.
\label{matterentropyspecificeos}
\end{equation}
The thermodynamic energy of the system given in 
Eq.~\eqref{thermalEtotal}
keeps the same expression, i.e.,
$E=R\left(1-\tilde{k}(R)\right)$, since it is independent
of the equation of state, and so is independent of
Eq.~\eqref{bhawktilder}.
Since the free energy $F$ of the system is $F=TI$ one finds
$F=R\left[1-\tilde{k}(R)\right]-T\pi \tilde{r}_+^2$, i.e., $F=E-TS$
as expected.

To analyze the thermodynamic stability we have calculate the heat
capacity $C_A$.  Putting the equation of state $b(\tilde{r}_+)=4\pi
\tilde{r}_+$, see Eq.~\eqref{bhawktilder}, into Eq.~\eqref{heatC}
yields
\begin{equation}
C_A= 4\pi R^2\left(\frac{\tilde{r}_+}{R}\right)^{2}
\frac{\left(1-\frac{
\tilde{r}_+}{R}\right)}{\left(3\frac{\tilde{r}_+}{R }
-2\right)}\,. \label{heatCex}
\end{equation}
Thus, from Eq.~\eqref{heatCstabil} the solution is stable, i.e.,
$C_A\geq0$, if the heat reservoir radius $R$ and the gravitational
radius $\tilde{r}_+$ obey the inequality
\begin{equation}
R\leq\frac32 \tilde{r}_+\,,
\label{R32tilder+}
\end{equation}
which is a condition also found by putting $b(\tilde{r}_+)=4\pi
\tilde{r}_+$, see Eq.~\eqref{bhawktilder}, into
Eq.~\eqref{Rstablecond}.

\subsection{Canonical ensemble solutions}

\subsubsection{Exterior solutions: Solutions
for the gravitational radius $\tilde{r}_+$ of the
hot thin shell}

Given the equation of state $b(\tilde{r}_+)=4\pi \tilde{r}_+$, see
Eq.~\eqref{bhawktilder}, one can calculate the possible
system's gravitational
radius solutions $\tilde{r}_+$.  Indeed, using
Eq.~\eqref{bhawktilder} in Eq.~\eqref{tolman2R} with
Eq.~\eqref{binverseT0} yields
$T=\frac{1}{4\pi\tilde{r}_+\tilde{k}(R)}$ which can be put in the
form
\begin{equation}
4\pi\frac{\tilde{r}_+}{R}
\sqrt{
1-\frac{\tilde{r}_+}{R}
}=\frac{1}{RT}\,.
\label{solutiontilder+}
\end{equation}
This equation, Eq.~\eqref{solutiontilder+},
is a cubic equation
for $\tilde{r}_+$.
For $0\leq RT<\frac{4\sqrt{27}}{32\pi}$
there are no solutions for it.
For $\frac{4\sqrt{27}}{32\pi}\leq
RT<\infty$
it has two solutions
for the gravitational radius of the system
as a function of $R$ and $T$. These solutions
can be written as
\begin{equation}
\tilde{r}_{+1}=\tilde{r}_{+1}(R,T)\,,
\label{tilder+1RT}
\end{equation}
\begin{equation}
\tilde{r}_{+2}=\tilde{r}_{+2}(R,T)\,,
\label{tilder+2RT}
\end{equation}
which we do not spell out explicitly, they are cubic root solutions of
the cubic equation given in Eq.~\eqref{solutiontilder+}.
The main thing of interest is that one solution, $\tilde{r}_{+1}$,
is small when compared to $R$, and the other
solution, $\tilde{r}_{+2}$, is
large.
 For
$RT=\frac{4\sqrt{27}}{32\pi}$ the two solutions merge in a degenerate
one.
The size
$\alpha$ of the shell is, in principle, independent of the
corresponding gravitational radius. Nonetheless, there is always the
constraint $\tilde{r}_+\leq\alpha$ and, as we will find soon,
$\alpha$ is further restricted, although not fixed, by other necessary
conditions on the solution.
Equations~\eqref{tilder+1RT}-\eqref{tilder+2RT} are the equivalent 
of the horizon radii of the two distinct
black holes of York's work
\cite{york1}, but now the equations
are for the gravitational radii of a system that contains a 
shell and has the specific equation of state given
above, Eq.~\eqref{bhawktilder}.
It interesting to comment on why for small $RT$, $0\leq
RT<\frac{4\sqrt{27}}{32\pi}$, there are no gravitational
radii solutions.  One
can associate to the temperature $T$ of the reservoir a wavelength,
the Compton thermal wavelength $\lambda_T=\frac{1}{T}$. So, small
$RT$, i.e., $RT$ less than about unity, specifically,
$RT<\frac{4\sqrt{27}}{32\pi}$, means small
$\frac{R}{\lambda_T}$, i.e., the thermal wavelength is greater than
the radius $R$ of the reservoir, and the associate radiation 
is stuck to the reservoir wall and cannot move freely.  For higher
$RT$, $RT$ greater than about unity, specifically,
$\frac{4\sqrt{27}}{32\pi}\leq RT<\infty$, then the associated thermal
wavelength is smaller than the radius $R$ of the reservoir, and
radiation can move freely and eventually form
structures such as a star in the form of a thin shell or a black
hole by direct collapse or tunneling.

There is yet another possible solution
for $\tilde{r}_+$, different from
$\tilde{r}_{+1}$ and $\tilde{r}_{+2}$
of Eqs.~\eqref{tilder+1RT} and \eqref{tilder+2RT},
respectively.
This is because
Eq.~\eqref{solutiontilder+} implicitly assumes that there is a shell
and a black hole inside the heat reservoir.  If there is no shell and
no black hole, the problem reduces itself to 
hot vacuum flat space thermodynamics.
In this case, 
there is flat space inside the reservoir, so
the gravitational radius of the system obeys
\begin{equation}
\tilde{r}_{+3}=0\,.
\label{newsolutiontilder+=0}
\end{equation}
This is the third possible solution
for $\tilde{r}_+$.
We note that the boundary for
the hot flat space solution
has the same topology as the
boundary for the black hole plus shell, namely,
$S^1\times S^2$.
This means, that although the spaces have
different topologies,
$S^1\times R^3$ in hot flat space,
$R^2\times S^2$ in black hole plus shell space,
they belong to the same canonical
ensemble, which is defined by the boundary data.
Thus, phase transitions between these
topological different spaces
are permitted, as it is known to be the case
in quantum, but not in classical, gravity.

\subsubsection{Interior Solutions: Solutions for the 
black hole horizon radius $r_+$}

Now, granted
that there are  solutions
$\tilde{r}_{+1}$ and $\tilde{r}_{+2}$,
Eqs.~\eqref{tilder+1RT} and \eqref{tilder+2RT},
we need to find the solutions for the horizon
radius $r_+$
of the assumed black hole inside the hot shell.
From Eq.~\eqref{rtdifferent} and with 
Eq.~\eqref{bhawktilder}
we have 
\begin{equation}
4\pi r_+\sqrt{1-\frac{r_+}{\alpha }}=4\pi
\tilde{r}_+\sqrt{1-\frac{
\tilde{r}_+}{\alpha }}\,.
\label{new1}
\end{equation}
Thus, the solutions for $r_+$ come from
Eq.~\eqref{new1}.
Note first
that by the same reasoning as
for Eq.~\eqref{solutiontilder+}
there are solutions $r_+$ if 
$\frac{4\sqrt{27}}{32\pi }\leq\frac{1}{4\pi
\frac{
\tilde{r}_+}{\alpha}\sqrt{1-
\frac{\tilde{r}_+}{\alpha }}}<\infty$.
For $0\leq\frac{\tilde{r}_+}{\alpha}\leq1$,
which is always the case, 
the two inequalities are always true, the
equality of the first inequality holding when
$\frac{
\tilde{r}_+
}
{\alpha}=\frac23$.
Making the substitution
$y=\frac{r_+}{\alpha}$
and
$z=\frac{\tilde{r}_+}{\alpha}$,
one obtains from 
Eq.~\eqref{new1} the equation 
$y\sqrt{1-y}=z\sqrt{1-z}$.
One solution of this equation is 
$y=z$ the other solution is
$y=\frac{1-z}{2}+\frac{\sqrt{(1-z)(1+3z)}}{2}$.
Let us analyze each solution in turn.

The first solution, $y=z$, is thus
\begin{equation}
r_{+1}=\tilde{r}_+\,,
\label{newsolutionr+}
\end{equation}
so this solution has no shell, it has only pure
black holes, actually two
distinct
black hole solutions, which incidentally are given by $\tilde{r}_{+1}$
or $\tilde{r}_{+2}$ of Eqs.~\eqref{tilder+1RT} and
\eqref{tilder+2RT}, respectively, and which have been discussed in
\cite{york1}.

The second solution,
$y=\frac{1-z}{2}+\frac{\sqrt{(1-z)(1+3z)}}{2}$,
is then
\begin{equation}
r_{+2}=\frac12\left(\alpha-\tilde{r}_+\right)
+\frac12
\sqrt{
\left(\alpha-\tilde{r}_+\right)\left(\alpha+3\tilde{r}_+\right)
}
\,.
\label{solutionr+}
\end{equation}
One can verify
that $r_{+2}<\alpha$ is automatically satisfied.  There
is also the necessary condition $0\leq r_{+2}\leq\tilde{r}_+$, where
$\tilde{r}_+$ can be either the gravitational radius
$\tilde{r}_{+1}$ or the gravitational radius $\tilde{r}_{+2}$ given in
Eqs.~\eqref{tilder+1RT} and \eqref{tilder+2RT}, respectively.  The
inequality of the necessary condition is obeyed for
$\frac{2}{3}\alpha\leq\tilde{r}_+\leq\alpha$, or
$
\tilde{r}_+\leq\alpha\leq\frac{3}{2}\tilde{r}_+
$,
where again $\tilde{r}_+$ can be either $\tilde{r}_{+1}$ or
$\tilde{r}_{+2}$ given in Eqs.~\eqref{tilder+1RT} and
\eqref{tilder+2RT}, respectively.  When the lower bound holds, i.e.,
$\tilde{r}_+=\frac{2}{3}\alpha$, one has $r_{+2}=\frac{2}{3}\alpha$,
i.e., $r_{+2}=\tilde{r}_+$, and one is back into
Eq.~\eqref{newsolutionr+}.  When the upper bound holds, i.e.,
$\tilde{r}_+=\alpha$, one has $r_{+2}=0$. The physical properties of
these solutions will be discussed soon.

There is yet another possible solution
for $r_+$, different from
$r_{+1}$ and $r_{+2}$
of Eqs.~\eqref{newsolutionr+} and \eqref{solutionr+},
respectively.
This is because Eq.~\eqref{new1}
implicitly assumes that there is a black hole inside the shell.  If
there is no black hole, the thermodynamics of the system with
a hot shell
alone has to have a different treatment. What matters here is that no
black hole, flat space inside the hot shell, is allowed, so
\begin{equation}
r_{+3}=0\,,
\label{newsolutionr+=0}
\end{equation}
is a third possible solution.

\subsubsection{Summary of the possible solutions}

The possible ensemble solutions for this system
with quantum properties,
comprised of a heat reservoir that can include
hot shells, black holes, and hot flat spaces,
that pop out from
the  equation of state
given in Eq.~\eqref{bhawktilder}
and from 
the action
Eq.~\eqref{actionspecific}, can be summarized by the following
combinations:
(i) no hot shell, which in turn has (a) $\tilde{r}_{+1}$, the small black
hole solution, (b) $\tilde{r}_{+2}$, the large black hole solution,
(c) $\tilde{r}_{+3}$, vacuum hot flat space, i.e., no black hole
solution, and (ii) hot shell solution with (a) $\tilde{r}_{+1}$, a hot
shell with small gravitational radius solution (1) with a black hole
inside with $r_{+2}$, or (2) with vacuum hot flat space inside,
$r_{+3}$, and (b) $\tilde{r}_{+2}$, a hot shell with large
gravitational radius solution (1) with a black hole inside with
$r_{+2}$, or (2) with vacuum hot flat space inside, $r_{+3}$.
So, there are seven possibilities, namely,
(i)(a), (i)(b), (i)(c), 
(ii)(a)(1),
(ii)(a)(2),
(ii)(b)(1), and 
(ii)(b)(2).

\subsubsection{Thermodynamic stability of the solutions}

In the case that $\frac{4\sqrt{27}}{32\pi} \leq RT<\infty$ we know
that there are two solutions for the hot shell, i.e., two
gravitational radius solutions, $\tilde{r}_{+1}(R,T)$, see
Eq.~\eqref{tilder+1RT}, and $\tilde{r}_{+2}(R,T)$, see
Eq.~\eqref{tilder+2RT}, the first shell with small $\tilde{r}_{+1}$
and so small gravitational mass, the second shell with large
$\tilde{r}_{+2}$ and so large gravitational mass.  We can now find the
thermodynamic stability of these two types of systems.

For that, we have to apply the heat capacity equation,
Eq.~\eqref{heatCex}, to the $\tilde{r}_{+1}$ and $\tilde{r}_{+2}$
solutions, and verify whether $C_A\geq0$, in brief, we have to verify
whether Eq.~\eqref{R32tilder+} is verified for each solution.  So the
solution $\tilde{r}_+$ is stable if Eq.~\eqref{R32tilder+} holds.
This means that the gravitational radius $\tilde{r}_{+1}(R,T)$
solution, see Eq.~\eqref{tilder+1RT}, being small, in fact
$R\geq\frac32 \tilde{r}_{+1}$, is unstable, and that the gravitational
radius $\tilde{r}_{+2}(R,T)$ solution, see Eq.~\eqref{tilder+2RT},
being large, in fact $R\leq\frac32 \tilde{r}_{+2}$, is stable.  One
can think that the unstable solution $\tilde{r}_{+1}$ can nucleate
into the stable $\tilde{r}_{+2}$ solution.
As we have already remarked, the
black hole inside with horizon radius $r_+$ has no direct
influence in the thermodynamic stability.

\subsection{Thermodynamic phases and phase transitions}

\subsubsection{Thermodynamic phases of the exterior solution:
Hot thin shells, pure black holes, and hot flat space}

In the canonical ensemble, for systems at fixed temperature, the free
energy $F$ is the important thermodynamic function.  Systems in
thermodynamic phases with higher $F$ tend to thermodynamic phases
where $F$ is a minimum. In the zero order approximation the classical
action $I$ is the energy $F$ up to the temperature of the system,
$F=TI$, and since $T$ is positive, we can make the analysis directly
in the action $I$.
Thus, in order to find which phases dominate when a hot
thin shell, a black hole, and hot flat space may be present we have to
compare the actions of the possible cases. Since a hot thin shell is a
curved space solution with matter, this system can give indications on
how the thermodynamic phases involving
black holes and hot matter in curved
spaces with quantum, i.e., semiclassical, properties 
behave between them.  All the analysis below, as
well as the previous analysis in this section~\ref{specificeos}, is
for the Hawking inverse temperature equation of state
$b(\tilde{r}_+)=4\pi \tilde{r}_+$, see Eq.~\eqref{bhawktilder}.

To start with, note that the action $I$ given in
Eq.~\eqref{actionspecific} depends on the ensemble fixed quantities
$R$ and $T$, and on the gravitational radius $\tilde{r}_+$
with $\tilde{r}_+=\tilde{r}_+(T,R)$. Thus
the action, and so the free energy, is independent of whether the
solution is a pure black hole with horizon radius $\tilde{r}_+$, as
it is implicit in Eq.~\eqref{newsolutionr+}, or the solution is a
shell with gravitational radius $\tilde{r}_+$ and with a black hole
inside, as it is implicit in Eq.~\eqref{solutionr+}, or the solution
is a shell with gravitational radius $\tilde{r}_+$ and with nothing
inside, as it is implicit in Eq.~\eqref{newsolutionr+=0}.  What
matters for the value of $I$ of Eq.~\eqref{actionspecific} is whether
one is evaluating it for $\tilde{r}_{+1}$ or for $\tilde{r}_{+2}$, see
Eqs.~\eqref{tilder+1RT} and \eqref{tilder+2RT}, respectively.  Thus,
$I(\tilde{r}_{+1})$ has the same value for the $\tilde{r}_{+1}$ pure
black hole, implicit in Eq.~\eqref{newsolutionr+}, for the
$\tilde{r}_{+1}$ shell with a black hole, implicit in
Eq.~\eqref{solutionr+}, or for the $\tilde{r}_{+1}$ shell with nothing
inside, implicit in Eq.~\eqref{newsolutionr+=0}.  As well,
$I(\tilde{r}_{+2})$ has the same value for the $\tilde{r}_{+2}$ pure
black hole, implicit in Eq.~\eqref{newsolutionr+}, for the
$\tilde{r}_{+2}$ shell with a black hole, implicit in
Eq.~\eqref{solutionr+}, or for the $\tilde{r}_{+2}$ shell with nothing
inside, implicit in Eq.~\eqref{newsolutionr+=0}.  Since what matters
for the value of the action $I$ of Eq.~\eqref{actionspecific} is
whether one is evaluating it at $\tilde{r}_{+1}$ or at
$\tilde{r}_{+2}$, let us analyze it at each of these two gravitational
radii.

For the solution $\tilde{r}_{+1}$, the small gravitational radius
solution, the action $I(\tilde{r}_{+1})$ has two features, it is a
maximum of $I(\tilde{r}_+)$ and it is greater than zero.  Since it is
a maximum, $\tilde{r}_{+1}$ gives an unstable solution, and since it
is greater than zero the solution either decays to hot vacuum flat
space which has zero action, $I=0$, or to the large black hole
solution $\tilde{r}_{+2}$ which has a value of the action lower than
$I(\tilde{r}_{+1})$ always, and in some cases lower than zero.  This
unstable $\tilde{r}_{+1}$ solution might be considered an instanton, a
solution that decays either to vacuum hot flat space or to the larger
gravitational radius solution $\tilde{r}_{+2}$. There is no phase with
the solution $\tilde{r}_{+1}$, and so it is of no interest for the
discussion of the possible phases.

For the solution $\tilde{r}_{+2}$, the large gravitational radius
solution, the action $I(\tilde{r}_{+2})$ is a minimum of
$I(\tilde{r}_+)$ and depending on the value of the product $RT$ it can
be greater than zero, equal to zero, or lower than zero.  Since it is
a minimum, it is a stable solution.  When it is lower than zero, which
happens for $RT$ sufficiently high, it is certainly a possible
thermodynamic phase of the system.

For the solution $\tilde{r}_{+3}$, there are no gravitational radii,
indeed the solution in this case is $\tilde{r}_{+3}=0$.  The solution
represents pure vacuum hot flat space and has action $I$ satisfying
$I(\tilde{r}_{+3})=0$.  Depending on the value of $RT$ it is a
possible thermodynamic phase of the system.

Let us spell out the details.  Since vacuum hot flat space
belongs to the ensemble and has zero action, $I=0$, one has now to
see how the action $I(\tilde{r}_{+2})$ changes with the boundary fixed
values $R$ and $T$, noting that $I(\tilde{r}_{+1})$ is of not interest
here since it corresponds to an unstable solution.  For $0\leq
RT<\frac{4\sqrt{27}}{32\pi}$ there are no gravitational radii
$\tilde{r}_+$, so no $\tilde{r}_{+2}$ solution, and the system
consists of vacuum hot flat space inside the reservoir.
For $\frac{4\sqrt{27}}{32\pi}\leq RT<\frac{27}{32\pi}$, there are two
gravitational radii $\tilde{r}_+$,
with $\tilde{r}_{+1}$ of no interest, and
with the $\tilde{r}_{+2}$
solution being a local minimum but yielding an action that is greater
than zero, $I(\tilde{r}_{+2})>0$, so that again vacuum hot flat space
having zero action, and thus zero free energy, is preferred.
For $\frac{27}{32\pi}\leq RT<\infty$, there are still two
gravitational radii $\tilde{r}_+$,
with $\tilde{r}_{+1}$ of no interest, and  with the $\tilde{r}_{+2}$
solution being a global minimum yielding an action that is equal or
lower than zero, $I(\tilde{r}_{+2})\leq0$, and so in this case the
large gravitational radius is the thermodynamic
phase that dominates the
ensemble. But now we note, that there are two possible
solutions for $\tilde{r}_{+2}$
of Eq.~\eqref{tilder+2RT}. One is
the pure black hole solution
of Eq.~\eqref{newsolutionr+}, $\tilde{r}_{+2}=r_{+1}$,
and the other is
the hot shell
solution $\tilde{r}_{+2}$
that splits into two possible solutions, 
one is $\tilde{r}_{+2}$ and $r_{+2}$,
i.e., the hot shell with a black hole
in its interior
of 
Eq.~\eqref{solutionr+},
and the other
is $\tilde{r}_{+2}$ and $r_{+3}$,
i.e., the hot shell with nothing
in its interior
of Eq.~\eqref{newsolutionr+=0},
For the same $\tilde{r}_{+2}$ all have the same action, i.e., 
$I_{\tilde{r}_{+2},r_{+1}}
=I_{\tilde{r}_{+2},r_{+2}\,{\rm or}\,r_{+3}}$, and 
so  pure black hole and the hot shell having the
same value of the action, a negative
value, have, at the semiclassical level, their dominant
phases coexisting in the ensemble, in a degenerate state.

\subsubsection{Thermodynamic
phases of the interior solution: Black hole and vacuum
hot flat space}

We now discuss the thermodynamic phases of the interior solution.
The
solutions with $\tilde{r}_{+1}$ are unstable and so of no interest to
the thermodynamic phases.  The solutions with $\tilde{r}_{+2}$ for
$\frac{4\sqrt{27}}{32\pi}\leq RT<\frac{27}{32\pi}$ are of no interest
in the phase analysis because vacuum hot flat space is preferable.
The solutions with $\tilde{r}_{+2}$ for $\frac{27}{32\pi}\leq
RT<\infty$ are of interest since they are thermodynamically favored,
and so we should discuss their interior, which can be given
by a pure black hole with horizon radius $\tilde{r}_{+2}$ of
Eq.~\eqref{newsolutionr+}, or by a shell with gravitational radius
$\tilde{r}_{+2}$ and with a black hole inside of
Eq.~\eqref{solutionr+}, or by a shell with gravitational radius
$\tilde{r}_{+2}$ and with nothing inside of
Eq.~\eqref{newsolutionr+=0}.
The pure black hole with horizon radius $\tilde{r}_{+2}$ of
Eq.~\eqref{newsolutionr+} has no interior by definition, so does
not enter into this analysis. 
It remains to analyze 
the shell with gravitational radius
$\tilde{r}_{+2}$ and with a black hole inside of radius $r_{+2}$
of 
Eq.~\eqref{solutionr+}, and the shell with gravitational radius
$\tilde{r}_{+2}$ and with nothing inside, i.e., $r_{+3}=0$, of
Eq.~\eqref{newsolutionr+=0}.
Since the action $I$ of
Eq.~\eqref{actionspecific} depends
on the gravitational radius $\tilde{r}_+$ of the
shell alone, we have
that 
$I_{\tilde{r}_{+2},r_{+2}}
=I_{\tilde{r}_{+2},r_{+3}}$, and 
so  the non existence or existence
of the black hole inside makes no
difference, the two interior phases coexist
at the semiclassical level in the ensemble,  
it is another degenerate state, this one for the interior.

\subsubsection{Possible phase transitions between the different
thermodynamic phases}

\begin{figure}[h]
\begin{center}
\includegraphics*[scale=0.75]{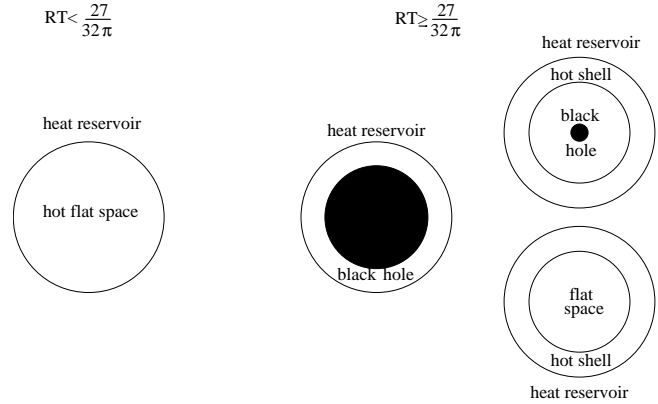}
\end{center}
\caption{The possible thermodynamic phases for
the canonical ensemble with a heat reservoir
with radius $R$ and at temperature $T$,
and for which 
the
system's reduced inverse temperature
equation of state is 
$b(\tilde{r}_+)=4\pi \tilde{r}_+$, are shown schematically.  The
phases in which the system settles depend
strongly on the reservoir values $R$ and $T$. For $0\leq
RT<\frac{27}{32\pi}$ the canonical ensemble systems settle in a hot
flat space phase.  For $\frac{27}{32\pi}\leq RT<\infty$ half of the
canonical ensemble systems settle in a pure black hole phase and half
in a hot thin shell phase, of which half have a black hole inside and
half have nothing, i.e., flat space, inside.
}
\label{phases}
\end{figure}

Let us summarize all the phases for
the equation of state 
$b(\tilde{r}_+)=4\pi \tilde{r}_+$,  collecting together the phases of the
exterior system solution, i.e., pure black holes,
hot shells, and hot flat
space, and the phases of the interior solution, i.e., black holes and
vacuum hot flat space, and make an analysis
 of the
possible phase transitions between the different
thermodynamic phases.
In Fig.~\ref{phases} the possible thermodynamic phases for
$b(\tilde{r}_+)=4\pi \tilde{r}_+$ are depicted for
the ensemble boundary quantities obeying
$0\leq RT<\frac{27}{32\pi}$ and
$\frac{27}{32\pi}\leq RT<\infty$.
The summary of the phases  is given after it.

\vskip 0.2cm
\noindent
$0\leq RT<\frac{27}{32\pi}$.  In this range we can split the analysis
into a first part given by $0\leq RT<\frac{4\sqrt{27}}{32\pi}$ and a
second part given by $\frac{4\sqrt{27}}{32\pi}\leq
RT<\frac{27}{32\pi}$.
For $0\leq RT<\frac{4\sqrt{27}}{32\pi}$
vacuum hot
flat space inside the reservoir is the direct phase. In this range of
$RT$ there are no minima, and actually no maxima as well, and so no
hot shells or black holes.  Indeed, here the associated Compton
thermal wavelength
$\lambda_T=\frac{1}{T}$
is greater than the radius $R$ of the reservoir,
and the radiation is locked at the reservoir wall,
cannot move freely, and hot shells and black holes cannot form.  Thus,
not considering some other quantum gravity phases that make no
appearance at the semiclassical level we are considering, only vacuum
hot flat space, a space with trivial topology, exists.  It is the
thermodynamic phase in which the system lives.
For $\frac{4\sqrt{27}}{32\pi}\leq RT<\frac{27}{32\pi}$,
there is a solution $\tilde{r}_{+2}$, be it a pure black hole or a hot
shell.  Indeed, the  associated Compton
thermal wavelength
$\lambda_T=\frac{1}{T}$ is smaller than the
radius $R$ of the reservoir, and the radiation can form
blob of matter in the form of
a shell or a black hole. However, the action $I$ of the
solution $\tilde{r}_{+2}$ here is greater than zero, i.e., the action
is greater than the action for vacuum hot flat space which has
zero action.
So $I(\tilde{r}_{+2})$ does not dominate the ensemble, instead vacuum
hot flat space with trivial topology is the ground state and is the
thermodynamic phase in which the system rests.  Thus, the two subcases
lead to hot flat space as the thermodynamic phase in which the system
settles.

\vskip 0.2cm
\noindent
$\frac{27}{32\pi}\leq RT<\infty$. In this range
there is a solution $\tilde{r}_{+2}$, be it a pure black hole or a hot
shell. Indeed, the   associated Compton
thermal wavelength
$\lambda_T=\frac{1}{T}$ is certainly
smaller than the
radius $R$ of the reservoir, and the radiation can form
blob of matter in the form of
a shell or a black hole.
Moreover, the action $I$ of the solution
$\tilde{r}_{+2}$ is a global minimum,
i.e., it is less than zero, and so it is less than the
action for vacuum hot flat space which has zero action.
Now, for the same
$\tilde{r}_{+2}$ the pure black hole and the hot shell have the same
value of the action, and so the two phases coexist in the ensemble.
In the case the phase is a hot shell there are two subphases, the
shell can contain a black hole or contain nothing, i.e., vacuum hot
flat space, and these two possible interior phases, having the same
action, coexist likewise.
In this $\frac{27}{32\pi}\leq RT<\infty$ case the
semiclassical approximation used is well justified, and so the result
for the various
quantum phases of the system in the canonical ensemble
is robust.

\subsection{The significant thermodynamic phase with its solutions and
additional comments}
\label{d}

We started with the mechanics of a black hole inside a thin shell and
displayed the essential properties of this static and spherical
symmetric general relativistic spacetime. Then, we added statistical
physics by setting the system in a canonical ensemble, i.e., the black
hole and the hot shell are enclosed in a heat reservoir at fixed
temperature and area.  We performed an initial analysis using the
Euclidean path integral approach to quantum gravity to calculate the
quantum partition function $Z$ from which the thermodynamics and other
statistical functions follow.  A zeroth order approximation was made
in that the classical action of the system enables one to find several
of the important quantum properties, at a semiclassical level, of
these microscopic gravitational systems.  Thus, the expression for the
gravitational action of the system was displayed, the entropy function
was found along with its consequences, the free energy and
thermodynamic energy were identified, distinct important features
related to the temperature and its stratification were shown, the
first law of thermodynamics was shown to hold, and the thermodynamic
stability was studied: Then, all of these generic features have been
applied to a system with a hot shell for which the temperature
equation of state is the Hawking temperature equation of state, an
equation of state that arises naturally from quantum fields in a black
hole space, namely, $b(\tilde{r}_+)= 4\pi\tilde{r}_+$.  Surprisingly,
the ensemble obliges to the presence of other solutions that are not
in the starting mechanical setting, i.e., the original shell with a
black hole inside.  The shell is now hot, the black hole is surely
hot, and the number of solutions in the ensemble increases. The
solutions that emerge are now a pure black hole, the original solution
of a black hole inside a hot shell, a solution of pure vacuum hot flat
space inside a hot shell, and the solution of only vacuum hot flat
space with no black hole and no shell. So, since there are two
possible solutions for the gravitational radius of the hot shell in
this ensemble, specifically, $\tilde{r}_{+1}$ and $\tilde{r}_{+2}$,
there are seven possible solutions instead of the initial one, of
which, when the thermodynamic phases are studied, only four remain,
the three related to $\tilde{r}_{+2}$ and the one of
$\tilde{r}_{+3}=0$.

Given these four solutions it is important to perform an analysis of
the thermodynamic phases that the system can assume and find out what
are the dominant ones.  In the analysis we have performed one can
single out the significant solutions and the corresponding phases,
where the semiclassical approximation for the Euclidean path integral
that we have followed is valid.  It is the solutions corresponding to
the exterior $\tilde{r}_{+2}$ solution, i.e., the solutions for the large
gravitational radius of the microscopic hot shell.  For this solution,
we showed that for the specific $b(\tilde{r}_{+2})=
4\pi\tilde{r}_{+2}$, one has $I(\tilde{r}_{+2})\leq0$, as long as the
ensemble quantities $R$ and $T$ obey $\frac{27}{32\pi}\leq RT<\infty$,
i.e., for sufficiently high $RT$. The corresponding solutions are
stable and the possible phases dominate the ensemble.  We have found
that a pure black hole and a hot thin shell coexist as phases in the
ensemble, and within the hot thin shell phase, the thin shell with
black hole inside coexists with the hot thin shell with nothing, i.e.,
vacuum hot flat space, inside.  Four comments are appropriate here.

First,
the system studied in a general setting, in particular the system with
equation of state $b(\tilde{r}_{+2})=4\pi\tilde{r}_{+2}$ that has been
analyzed in detail, is of interest when the temperatures of its
components are non-negligible and have a real influence in the
thermodynamic and thermal properties of the system.  This means that
the heat reservoir, the thin shell, and the black hole have to be of
microscopic size, where the curvature of space is sufficiently high
such that the associated quantum effects imply a significant emission
of radiation from the black holes and the other components of the
system that together are in an equilibrium thermodynamic state.

Second,
in this microscopic, semiclassical, 
setting for the canonical
ensemble of specific gravitational
systems, we are able to compare a black hole phase with a
curved space hot matter phase, here realized as a hot thin shell, in
the canonical ensemble.  This is of real interest because since black
holes have mass and the corresponding space is curved, one
expects that when they fully evaporate, some other curved space filled
with hot matter might arise.  This hot matter could be composed of hot
gravitons, hot photons, or other hot particles. We have reproduced the
hot matter in a hot thin shell, which is a system amenable to an exact
thermodynamic solution as we have showed.

Third,
throughout the whole analysis, in particular for the analysis of the
phases, it has become clear that the inverse temperature equation of
state $b(\tilde{r}_+)$ is overwhelmingly important.  This is because
the system's temperature equation of state, represented in
$b(\tilde{r}_+)$, dictates whether black hole phases dominate over hot
shell phases or not, or whether both phases coexist in the canonical
ensemble.  In general, a specific inverse temperature equation of
state might give that one phase must dominates over the other, and, in
the case the hot thin shell dominates, it must state whether a hot
thin shell with a black hole inside phase dominates over a hot thin
shell with nothing inside.  For the specific, particular, choice of
$b(\tilde{r}_+)=4\pi \tilde{r}_+$, we have seen that the pure black
hole and the hot thin shell phases coexist in the ensemble. This
result makes sense, a black hole and a hot shell that have the same
form of the temperature equation of state must exist side by side in a
thermodynamic setting, thermodynamically the two phases cannot be
distinguished.  Moreover, in the hot thin shell phase, we have also
seen that for $b(\tilde{r}_+)=4\pi \tilde{r}_+$, the hot thin shell
with a black hole inside phase coexists with a hot thin shell with
nothing inside. For an analysis of the phases
for other equations of state see Appendix \ref{othereos}.

Fourth, connected to the previous comment, we can say that for
$b(\tilde{r}_+)=4\pi \tilde{r}_+$, thermodynamic information
to the exterior of the reservoir
about what is in the inside of the boundary at $R$
is not explicit. A
microscopic
observer perching at the reservoir, interested in studying the
thermodynamics of the microscopic gravitational
system, cannot distinguish thermodynamically the
various phases that the system might be in.  This is a kind of black
hole thermodynamic mimicker, rather than black hole dynamical
mimicker.  Moreover, in the particular case that the gravitational
radius of the hot shell $\tilde{r}_+$ is close to its radius $\alpha$,
then the configuration itself is a quasiblack hole, and so in this
very case one is in the presence of a thermodynamic plus dynamical
and geometric black hole mimicker.

\section{Analytical solutions in the high temperature limit for the
reservoir, the hot self-gravitating
thin shell, and the black hole in the canonical
ensemble} \label{highT}

\subsection{Finite temperature heat reservoir and high temperature
shell, two cases}

\subsubsection{The case of the hot shell radius, the hot shell
gravitational radius, and the black hole horizon radius nearly equal,
i.e., $\tilde{r}_+\to\alpha$ and $r_+\to\alpha$: Analytical solution}

We now study the very interesting case
of finite heat reservoir
temperature $T$, which together with the radius $R$
is kept fixed as it should in the canonical
ensemble,
and very high, tending to infinity,
shell temperature $T(\alpha)$,
with $\tilde{r}_+\to\alpha$
and 
$r_+\to\alpha$, recalling that
$r_+\leq \tilde{r}_+\leq\alpha$.
In brief, one has in this high temperature case
\begin{equation}
T={\rm finite}, \,T(\alpha)\to\infty,\;\;
{\rm with}\,\,\,\,
\tilde{r}_+\to\alpha,\,\,
r_+\to\alpha.
\label{casea1}
\end{equation}
The aim is to find $\alpha$ and $r_+$
in this case as a function of
the relevant parameters.
We will see
that in this high temperature limit we can determine
$\alpha$ and $r_+$ in terms of a generic
equation of state for the function $b$.
Then if we want to determine 
the system thermodynamically, a specific equation of state
$b$ has to be given,
with  $\tilde{r}_+$
being then fixed by the equation of state
$b(\tilde{r}_+)$, once this is given,
through Eq.~\eqref{tolman2R}, or  Eq.~\eqref{binverseT0}.
Before embarking on the calculation of $\alpha$ and $r_+$ in this high
temperature limit we make a preamble on notation for what follows.
Note that the redshift factor $\tilde{k}$ at $R$ is written as
$\tilde{k}(R)$ with $R$ fixed, but it can be envisaged as
$\tilde{k}(R,\tilde{r}_+)$, since from Eq.~\eqref{redstilder} one
has $ \tilde{k}=\tilde{k}(R,\tilde{r}_+)=
\sqrt{1-\frac{\tilde{r}_+}{R}}$.
As well, one has that the
temperature at an inner $r$, $T(r)$, in Eq.~\eqref{localTfromi},
 is
actually
$T(r,r_+)= \frac{1}{4\pi r_+k(r,r_+)} $, with
$k(r,r_+)=\sqrt{1-\frac{r_+}{r}}$.
One can say
the same for the
temperature at the shell $T(\alpha)$, in Eq.~\eqref{Talpha1}, it is
actually $T(\alpha,r_+)= \frac{1}{4\pi r_+k(\alpha,r_+)} $, with
$k(\alpha,r_+)=\sqrt{1-\frac{r_+}{\alpha}}=k(\alpha)$.  
And surely, the reservoir
temperature $T$, fixed in the ensemble, is sometimes written as $T(R)$
with $R$ also fixed in the ensemble, but it can be envisaged as
$T(R,\tilde{r}_+)$, since from Eq.~\eqref{binverseT0} one has
$T=T(R,\tilde{r}_+)=
\frac{T_{0}(\tilde{r}_+)}{\tilde{k}(R,\tilde{r}_+)} =
\frac{T_{0}(\tilde{r}_+)} {\sqrt{1-\frac{\tilde{r}_+}{R}} }$.
On the other
hand, since $\tilde{k}$ is a function of two variables,
it can happen that
the second variable gets the value $\alpha$,
so we have in this case
$ \tilde{k}(R,\tilde{r}_+=\alpha)=\sqrt{1-
\frac{\alpha }{R}}$.
To shorten the notation
we write
$\tilde{k}_\alpha\equiv \tilde{k}(R,\tilde{r}_+=\alpha)=\sqrt{1-
\frac{\alpha }{R}}$.
But $\tilde{k}_\alpha$ is the same as 
$k_\alpha$, since $k_\alpha\equiv k(R,r_+=\alpha)=
\sqrt{1-\frac{\alpha }{R}}$, and we
stick to 
$k_\alpha$.
Also, since $T$ is a function of two variables, it can happen that
the second variable gets the value $\alpha$,
so we have in this case
that $T_\alpha\equiv T(R,\tilde{r}_+=\alpha)
=T(R,\alpha)$,
with  $T_\alpha$
written to shorten the notation.
Note that $T_\alpha$ is different from $T(\alpha)$,
this latter happens when we substitute the
first argument $r$ for $\alpha$,
$T(\alpha)=
T(r=\alpha,r_+)
=T(\alpha,r_+)$,
in words, $T_\alpha$ is still
the local temperature on the boundary $R$ 
but the gravitational radius $\tilde{r}_+$
is swapped with $\alpha$, whereas
$T(\alpha)$ is the local temperature on the shell.  Note that since
$b$ is a function of $\tilde{r}_+$ alone, $b=b(\tilde{r}_+)$, one has
$b_\alpha=b(\alpha)$.  All this notation is of relevance now.

Let us, maintaining $T$ finite, take the high temperature limit
for the shell, i.e., 
$T(\alpha )\rightarrow \infty $. We see from
Eq.~\eqref{tolmanoutalpha} that this means 
$\tilde{r}_+\rightarrow \alpha$,
and from 
Eq.~\eqref{Talpha1} one of the possibilities
is that $r_+\rightarrow \alpha$ also.
Recall that $r_+\leq \tilde{r}_+$.
Then, taking into account the notation
of the preamble of this section, we can expand
Eq.~\eqref{binverseT0} for $\tilde{r}_+$ near $\alpha$ as
$T(R,\tilde{r}_+)= T(R,\tilde{r}_+)\vert_{\tilde{r}_+=\alpha}
+
{\frac{\partial T(R,\tilde{r}_+)}{\partial
\tilde{r}_+}}\vert_{\tilde{r}_+=\alpha}
(\tilde{r}_+-\alpha )$
plus higher order terms in $\tilde{r}_+-\alpha$.
Now, $T(R,\tilde{r}_+)\vert_{\tilde{r}_+=\alpha}
=T(R,\tilde{r}_+=\alpha)
=T(R,\alpha)\equiv T_\alpha$
as defined above. Since from Eq.~\eqref{binverseT0},
see also Eq.~\eqref{tolman2R},
one has 
$T(R,\alpha)=
\frac{1}{b(\alpha)\tilde{k}(R,\alpha)}$,
then, with the definitions above, we can write
\begin{equation}
T_{\alpha }=\frac{1}{
b_\alpha {k}_\alpha}\,,
\label{Tal}
\end{equation}
where we have used $\tilde{k}_{\alpha}={k}_{\alpha}$.
Note that, in this limit,
$T_\alpha$ is finite
and $T(\alpha)$ is going to infinite,
so clearly they are different quantities.
We also have to find an appropriate expression for 
${\frac{\partial T(R,\tilde{r}_+)}{\partial
\tilde{r}_+}}\vert_{\tilde{r}_+=\alpha}$
in this limit.
From the definition of $C_A$ one has
$C_A=\left(\frac{\partial E}{\partial T}\right)_A$,
so 
$C_A=
\left(\frac{\partial E}{ \partial
\tilde{r}_+}\right)_A
\frac{\partial \tilde{r}_+}{\partial T}$.
Inverting, we get $
\frac{\partial T}{\partial  \tilde{r}_+}=
\frac{\left(\frac{\partial E}{\partial
\tilde{r}_+}\right)_A}
{C_A}$.
Since $E=R\left(1
-\tilde{k}\right)$, see Eq.~\eqref{thermalEtotal}, we have
$\left(\frac{\partial E}{\; \partial
\tilde{r}_+}\right)_A=-R
\left(\frac{\partial \tilde{k}}{\; \partial
\tilde{r}_+}\right)_A
$
and since 
$\tilde{k}=\sqrt{1-\frac{\tilde{r}_+}{R}}$ 
we have 
$\left(\frac{\partial \tilde{k}}{\; \partial
\tilde{r}_+}\right)_A=-\frac{1}{2R\tilde{k}}$,
so that 
${\frac{\partial T(R,\tilde{r}_+)}{\partial
\tilde{r}_+}}\vert_{\tilde{r}_+=\alpha}=\frac1{2{k}_\alpha C_\alpha}$,
plus higher order
terms, and where $C_\alpha$ is 
$C_A$ evaluated at $\tilde{r}_+=\alpha$ keeping
our convention.
Then, from Eq.~\eqref{heatC} one has
\begin{equation}
C_\alpha=\frac{b^2_\alpha  {k}_\alpha^2R}
{b_\alpha-2b^{\prime}_\alpha  {k}_\alpha^2
R}\,,
\label{Calpha}
\end{equation}
where again we have used $\tilde{k}_{\alpha}={k}_{\alpha}$.
Then, the expansion
$T(R,\tilde{r}_+)= T(R,\tilde{r}_+)\vert_{\tilde{r}_+=\alpha}
+
{\frac{\partial T(R,\tilde{r}_+)}{\partial
\tilde{r}_+}}\vert_{\tilde{r}_+=\alpha}
(\tilde{r}_+-\alpha )$ plus higher order terms can be written
as
\begin{equation}
T= T_\alpha+
\frac{1}{2C_\alpha{k}_\alpha
}
(\tilde{r}_+-\alpha )\,,
\label{Tnear}
\end{equation}
plus higher order terms, 
where $T_\alpha$ is given in Eq.~\eqref{Tal} and $C_\alpha$ is given
in Eq.~\eqref{Calpha}, and we have now omitted the arguments of $T$ to
reinforce that $T$ is fixed in the ensemble,
as well as $R$. 
The condition $\tilde{r}_+<\alpha$ means from Eq.~\eqref{Tnear}
that, since ${k}_\alpha=\sqrt{1-\frac{\alpha}{R}}>0$,
one has
$C_\alpha
\left(T-T_{\alpha}\right)<0$. Then from Eq.~\eqref{Tnear}
one can write
\begin{equation}
\tilde{r}_+= \alpha -
2{k}_\alpha\left\vert C_\alpha (T-T_{\alpha})\right\vert
\,,
\label{rtildehigh}
\end{equation}
plus higher order terms.  Note that Eq.~\eqref{rtildehigh} is
deceiving. The reason is that as it is written it seems an equation
for $\tilde{r}_+$, but is in fact an implicit equation for $\alpha$,
i.e., $\alpha=\alpha(T,R)$. The reason, already referred, is that
$\tilde{r}_+$ is determined from $T$ and $R$ from the start,
$\tilde{r}_+=\tilde{r}_+(T,R)$ as seen in Eq.~\eqref{binverseT0}, see
also Eq.~\eqref{tolman2R}.  Once one knows the solution for
$\tilde{r}_+$ then, in this limit, $\alpha$ can be determined
afterward, it is given implicitly by Eq.~\eqref{rtildehigh}.  Thus, in
this high temperature regime, $\alpha$ is determined, a thing that
does not happen generically, indeed, in general $\alpha$ belongs to
the data of the problem.  To know, the solutions for $\tilde{r}_+$ one
needs to give the equation of state $b(\tilde{r}_+)$, but the whole
analysis in this high temperature case can be carried out without
specifying an equation of state up to the end.  Note that we can also
envisage Eq.~\eqref{rtildehigh} in the following way: We can put
Eq.~\eqref{rtildehigh} as an equation of the form $T=T(R,\alpha)$.
Then, we can fix $\alpha$ and, for each $\alpha$, find the curve
$T=T(R)$, which is a curve for the ensemble boundary data, along
which, in the space of parameters, we have the high temperature limit
under discussion.

Another interesting result in the high temperature regime is to find
the inner black hole
horizon radius
$r_+$, which should be lower than $\tilde{r}_+$ by a tiny amount.
From Eq.~\eqref{rt}, i.e.,
$4\pi r_+k(\alpha)=
b(\tilde{r}_+) \tilde{k}(\alpha)$ or in full
$4\pi r_+\sqrt{1-\frac{r_+}{\alpha }}=
b(\tilde{r}_+) \sqrt{1-\frac{ \tilde{r}_+}{\alpha }}$, one finds
$b^2(\tilde{r}_+)\left(\alpha-\tilde{r}_+\right) =16\pi^2
r_+^2\left(\alpha-r_+\right) $. Noticing further that
$\alpha-\tilde{r}_+$ and $\alpha-r_+$ are already first order,
one can put $b^2(\tilde{r}_+)=b^2(\alpha)$ and $16\pi^2
r_+^2=16\pi^2\alpha^2$, and the equation turns into 
$b^2(\alpha)\left(\alpha-\tilde{r}_+\right) =16\pi^2\alpha^2
\left(\alpha-r_+\right)$. Using Eq.~\eqref{rtildehigh} one
then gets
\begin{equation}
r_+= \alpha -\frac{b^2_\alpha}{16\pi ^2\alpha ^2}
2k_\alpha\left\vert C_\alpha (T-T_{\alpha})\right\vert
\label{r+highT}\,,
\end{equation}
plus higher order terms, and where, following our notation,
we have put
$b^2(\alpha)= b^2_\alpha$.
Note that Eq.~\eqref{r+highT}
is an equation for $r_+$, i.e., $r_+=r_+(T,R,\alpha)$,
see also Eq.~\eqref{rt}
and the  discussion after it,
where $T$ and $R$ are fixed by the ensemble,
and $\alpha$ is determined implicitly in 
Eq.~\eqref{rtildehigh},
so that in the end, in this
high temperature limit 
$r_+=r_+(T,R)$.
To finalize the solution
one needs to give the equation of
state $b(\tilde{r}_+)$, but as we see from
Eq.~\eqref{r+highT}, $r_+$ can be determined
generically for generic $b$.

Further comments are given now.
First,
the condition $r_+\leq \tilde{r}_+$ implies
with the help of Eqs.~\eqref{rtildehigh}
and~\eqref{r+highT} that
$b_\alpha>4\pi \alpha$,
or if one prefers $b(\alpha)>4\pi \alpha$,
which was also found generically from Eq.~\eqref{rt}.
Note that from Eq.~\eqref{junctionmass}
one finds that in this case
the rest mass $m$ of the thin shell is
$m=\left(\frac{b_\alpha}{4\pi\alpha}-1\right)
\sqrt{2
\alpha k_\alpha
\left\vert C_\alpha (T-T_{\alpha})\right\vert}$,
which is tiny, since
$\left\vert T-T_{\alpha} \right\vert$
is tiny.
Second, it is interesting
to calculate $T(\alpha)$ in this limit.
From
the Tolman relation $T(\alpha) \tilde{k}(\alpha)=
T(R) \tilde{k}(R)$ and 
Eq.~\eqref{binverseT0}
one has $T(\alpha )=\frac{T_{0}(\tilde{r}_+)}{\tilde{k}(\alpha )}$,
or
$T(\alpha)=\frac1{b(\tilde{r}_+)
\sqrt{1-\frac{\tilde{r}_+}{\alpha}}}$, which upon substitution of
Eq.~\eqref{rtildehigh}, yields
\begin{equation}
T(\alpha)=
{T_\alpha}
\sqrt{
\frac{\alpha}{2{k}_\alpha
\left\vert C_\alpha (T-T_{\alpha})\right\vert}}\,,
\end{equation}
plus higher order terms,
which is high since $\left\vert T-T_{\alpha}\right\vert$ is small.
Third, since $r_+$ tends to 
$\tilde{r}_+$ which tends to $\alpha$,
one could think that the configuration
is tending to a quasiblack hole with a large
black hole inside the shell, but this is not the
case because the mass of the shell
is tending to zero, leaving in the limit a black
hole alone with infinite temperature at
its own horizon as it should.
Fourth, this microscopic
gravitational system can be thermodynamically stable as long as 
$
R\leq \tilde{r}_++
\frac{b(\tilde{r}_+)}{2b^{\prime}(\tilde{r}_+)}$, see
Eq.~\eqref{Rstablecond}, so
in this limit one has $
R\leq \alpha+
\frac{b_\alpha}{2b^{\prime}_\alpha}$.
Fifth, note that this high temperature case
cannot be realized by the equation
of state Eq.~\eqref{bhawktilder}
of the previous section, Sec.~\ref{specificeos},
since this case requires $b(\alpha)>4\pi \alpha$,
and the equation
of state Eq.~\eqref{bhawktilder} yields 
$b(\alpha)=4\pi \alpha$.

\subsubsection{The case of the hot shell radius
nearly equal to the hot shell gravitational radius
and the black hole horizon radius going to zero, i.e.,
$\tilde{r}_+\to\alpha$ and $r_+\to0$}

We now study the other case of finite heat reservoir temperature $T$
which together with $R$ is kept fixed, 
and very high, tending to infinity, shell temperature $T(\alpha)$, now
with $\tilde{r}_+\to\alpha$ and $r_+\to0$. In brief, one has
in this high temperature case
\begin{equation}
T={\rm finite}, \,T(\alpha)\to\infty,\;\; {\rm with}\,\,\,\,
\tilde{r}_+\to\alpha,\,\, r_+\to0.
\label{casea2}
\end{equation}

Indeed, maintaining $T$ finite, and taking the high temperature limit
for the shell, i.e., $T(\alpha )\rightarrow \infty $, we see from
Eq.~\eqref{tolmanoutalpha} that this
again means $\tilde{r}_+\rightarrow
\alpha$, and from Eq.~\eqref{Talpha1} there is
a further possibility, namely, 
$r_+\rightarrow 0$.  The configuration is tending to a quasiblack hole
with a small black hole inside the hot shell,
and thus the system acts as a dynamical and geometric
black hole mimicker.
This microscopic
gravitational system can be thermodynamically stable as long as 
$
R\leq \tilde{r}_++
\frac{b(\tilde{r}_+)}{2b^{\prime}(\tilde{r}_+)}$, see
Eq.~\eqref{Rstablecond},
i.e.,
$
R\leq \alpha+
\frac{b(\alpha)}{2b^{\prime}(\alpha)}$.
This infinite temperature
case is realizable by the equation of state $b(\tilde{r}_+)=4\pi
\tilde{r}_+$ of Sec.~\ref{specificeos}, specifically,
Eq.~\eqref{bhawktilder}.

\subsection{High temperature heat reservoir and high temperature
shell, two cases}

\subsubsection{The case
of the hot shell gravitational radius going to $\alpha$ going
to $R$ and the black hole horizon radius going to zero, i.e.,
$\tilde{r}_+\to\alpha\to R$ and $r_+\to0$}

We now study
divergingly high
heat reservoir temperature $T$
and divergingly high shell temperature
$T(\alpha)$, with $\tilde{r}_+\to\alpha\to R$ and $r_+\to0$.
In brief, one has in this
high temperature case
\begin{equation}
T\to\infty, \,\,T(\alpha)\to\infty,\;\; {\rm with}\,\,\,\,
\tilde{r}_+\to\alpha\to R,\,\, r_+\to0.
\label{case2a}
\end{equation}

Indeed, if the heat reservoir temperature $T$ is very
high then the shell temperature $T(\alpha)$ is also very high,
and from
Eq.~\eqref{binverseT0},
or Eq.~\eqref{tolman2R},
one finds the case
$\tilde{r}_+\to\alpha\to R$ and
$r_+\to0$.
This means the shell has
high mass, and so this is essentially York's large
black hole solution when it
goes to radius $R$, but now it is a
large quasiblack hole solution.
Thus the system  acts as a dynamical and geometric
black hole mimicker.
For any $b(\tilde{r}_+)$ such that
$b'(\tilde{r}_+)>0$,
this microscopic
gravitational system is thermodynamically stable, since 
$
R\leq \tilde{r}_++
\frac{b(\tilde{r}_+)}{2b^{\prime}(\tilde{r}_+)}$, see
Eq.~\eqref{Rstablecond},
turns into $
R\leq R+
\frac{b()}{2b^{\prime}(R)}$,
a condition that is obeyed.
This infinite temperature
case is realizable by the equation of state $b(\tilde{r}_+)=4\pi
\tilde{r}_+$
of Sec.~\ref{specificeos}, specifically,
Eq.~\eqref{bhawktilder}.

\subsubsection{The case of the shell gravitational radius going to
zero and the black hole horizon radius going to zero, i.e.,
$\tilde{r}_+\to0$ and $r_+\to0$}

We study yet another case,  of
heat reservoir with divergingly high
temperature $T$ and divergingly high shell temperature
$T(\alpha)$, with $\tilde{r}_+\to0$ and $r_+\to0$,
recalling that $r_+\leq \tilde{r}_+\leq\alpha$.
In brief, one has in this
high temperature case
\begin{equation}
T\to\infty, \,\,T(\alpha)\to\infty,\;\; {\rm with}\,\,\,\,
\tilde{r}_+\to0,\,\, r_+\to0.
\label{case2b}
\end{equation}

Indeed, if the heat reservoir temperature $T$ is very high then there
is another situation.  It comes
from
Eq.~\eqref{binverseT0}, 
or
Eq.~\eqref{tolman2R}, when
one makes $b(\tilde{r}_+)\to0$,
which yields
$b(\tilde{r}_+)=4\pi \gamma \tilde{r}_+$
for some $\gamma<1$, and then
from Eq.~\eqref{rtdifferent} one has $r_+=\gamma \tilde{r}_+$,
yielding additionally that $r_+\to0$.
This means that the shell has
very small mass, and so this is essentially York's small black hole
solution when it goes to zero. Since in the exterior there is a shell,
with small mass but not necessarily comparatively small in size, it
can pass as a mimicker for a small black hole.  This microscopic
gravitational system is thermodynamically unstable, since $ R\leq
\tilde{r}_++ \frac{b(\tilde{r}_+)}{2b^{\prime}(\tilde{r}_+)}$, see
Eq.~\eqref{Rstablecond}, i.e.,
$R\leq 0$ is certainly a condition that is not obeyed.  This
infinite temperature case is realizable by the equation of state
$b(\tilde{r}_+)=4\pi \tilde{r}_+$ of Sec.~\ref{specificeos},
specifically, Eq.~\eqref{bhawktilder}.

\section{Conclusions}
\label{conc}

We have studied gravitational systems in the canonical ensemble that
provide a deeper insight to the connection between black holes and hot
matter fields in curved spaces and allow for the possibility of the
conversion of the first into the second and vice versa, giving thus
a further contribution to the understanding of the Euclidean path
integral approach to quantum gravity.

Indeed, the result that in the canonical ensemble the entropy of
systems that include matter, as examples of non pure black hole
systems, depends only on the gravitational radius ${\tilde r}_+$ of
the matter is an important result. The entropy can then be written as
$S=S(\tilde{r}_+)$ or as $S=S(\tilde{A}_+)$,
with the area corresponding to the gravitational
radius being given by $\tilde{A}_+=4\pi\tilde{r}_+^2$.
As well, the conclusion that the black hole inside the matter has no
direct influence on the entropy of the system is a noteworthy
development of the formalism used.
Moreover, the finding that, in the
semiclassical approximation, the canonical ensemble of systems with
matter presents several possible phases, including shell and black
hole phases, as was shown within a dynamic and thermodynamic
exact solution, is a specific consequence of the approach which is
displayed here  for the first time.
Similarly, the circumstance of having phase transitions between spaces with
different topologies, namely, black hole spaces, shell spaces, and hot
flat spaces, is highly interesting, as it is another instance where
quantum gravity effects can realize topological phase transitions
between spaces. Actually, the result that in the canonical ensemble new
prospects at the semiclassical level emerge through the finding of
nontrivial exact solutions of the thermodynamic systems with different
topologies is remarkable, developing thus further ideas on this issue
that started some time ago.
In addition, the idea put forward in our work that there are thermodynamic
black hole mimickers, besides the well-known dynamical
mimickers, is new. 
Still, the analytic solution found by us in the high temperature limit is
of note, not the least because several conclusions can be taken
without specifying an equation of state.

The canonical ensemble together with the Euclidean path integral
approach to quantum gravity in the semiclassical approximation
that we have
been considering is of interest for gravitational systems microscopic
in size, where quantum effects are important and the Hawking
temperature of the black hole is significant and has an impact on the
system.  Here, the system studied provided an instance where the
canonical ensemble is constituted of a mixture of quantum black holes
and curved space with hot matter, realized here as a hot thin
shell. It showed the new possibilities for the semiclassical
treatment of more complex systems in the canonical ensemble, and
has indicated how one may enlarge the notion of black hole
mimickers to incorporate systems that are not black holes but
can mimic their thermodynamic behavior.

\begin{acknowledgments}
We thank Tiago Fernandes for conversations on black holes and
hot thin shells in the Euclidean path integral approach.
We acknowledge financial support from Funda\c c\~ao para a Ci\^encia e
Tecnologia - FCT through the project~No.~UIDB/00099/2020.
\end{acknowledgments}

\appendix

\section{Generic inverse temperature 
equations of state $b(\tilde{r}_+)$: Some remarks}
\label{othereos}

\subsection{Preliminaries}

The function $b(\tilde{r}_+)$ is the important function in the
thermodynamic study of a hot thin shell with a
black hole inside in the
semiclassical approximation for the canonical ensemble.  We have
assumed in Sec.~\ref{specificeos} that the inverse
temperature equation of state for the hot shell is
$b(\tilde{r}_+)=4\pi \tilde{r}_+$, i.e., the Hawking equation of
state. But of course one can assume many other reduced inverse
temperature equations of state, as long as one chooses a
$b(\tilde{r}_+)$ that is physically reasonable, specifically, not
singular or not violating physical principles.

In the form of an example, We have suggested
a power law equation of state for
$b(\tilde{r}_+)$, $b(\tilde{r}_+)={\gamma \ell}\left(
\frac{\tilde{r}_+}{\ell} \right)^{\hskip -0.1cm\delta}$, for some
$\gamma$, $\ell$, and $\delta$, obeying
$0\leq\gamma<\infty$, $0<\ell<\infty$,
and $0\leq\delta<\infty$.  The case $\delta$ very
small is interesting.  Putting $\delta\to0$ one has for the heat
capacity that $C_A= \gamma \ell^2 \left(1-\frac{
\tilde{r}_+}{R}\right)$, see
Eqs.~\eqref{heatC} and \eqref{Rstablecond}.
We thus deduce that any reservoir radius $R$,
$1\leq\frac{R}{\tilde{r}_+}<\infty$, gives a thermodynamically
stable solution.  So, in this case of small $\delta$, the heat
reservoir can be put as far as one wants from the gravitational radius
and from the photonic orbit radius of the shell, as well as
from the radius 
$\alpha$ of the shell itself. This
is in contrast to the case $\delta=1$ of Sec.~\ref{specificeos},
where $R$ has to be at the photonic radius or inside it to have a
system that is
thermodynamically stable with the consequence
that within that region
dynamic instabilities can set in. In the case of small $\delta$
these
possible dynamic instabilities do not appear as
the heat reservoir radius is as large as one wants.

Another example for a specific
$b(\tilde{r}_+)$ would be a polynomial with several different powers
in $\tilde{r}_+$ which could give more solutions for $\tilde{r}_+$
than the two we have found for $b(\tilde{r}_+)=4\pi\tilde{r}_+$.
Then, depending on the form of the specific polynomial
one can have several solutions that
are thermodynamically stable as well as others that are unstable.

From these two examples for $b(\tilde{r}_+)$, one can see that
the specific inverse temperature equation of state $b(\tilde{r}_+)$
of the shell plays an essential role in the thermodynamic and dynamic
behavior of the system.
Here we do not wish to make an exhaustive study of possible equations
of state for $b(\tilde{r}_+)$.  Rather, we want to understand how
$b(\tilde{r}_+)$ influences the phases of the canonical ensemble.
To keep the analysis generic we do not explicitly indicate an
expression for $b(\tilde{r}_+)$. In this way, we will be able to
make some important remarks when performing a
comparison of the actions of the different possible thermodynamic
phase solutions leaving $b(\tilde{r}_+)$ unspecified.

\subsection{Dominant phases}

\subsubsection{Hot thin shell with black hole inside versus hot thin
shell with nothing inside: Dominant thermodynamic phases}

Let us start by calculating the difference between the action for a
hot thin shell with black hole inside and the action for a hot thin
shell with nothing inside. We are assuming that the hot thin shell
action dominates the ensemble, see
Eq.~\eqref{action1}.  So, still with $b(\tilde{r}_+)$ generic, we
want to calculate $I_{\rm hot \, shell\, with \,bh} -
I_{\rm hot \, shell\, with\, flat \,space}$.  We have from
the action given in Eq.~\eqref{action1}
and with the help of Eq.~\eqref{matterentropy}
that $I_{\rm hot \, shell\, with \,bh}=
\beta R\left[1-\tilde{k}(R)\right]-\pi r_+^2
-\frac12\int_{r_+}^{\tilde{r}_+}b(r)\,dr$.  We also have from
Eq.~\eqref{action1} that $I_{\rm hot \, shell\, with\, flat \,space}=
\beta R\left[1-\tilde{k}(R)\right] -\frac12
\int_0^{\tilde{r}_+}b(r)\,dr$, since here $r_+=0$.
So,
\begin{align}
I_{\rm hot \, shell\, with \,bh} -& I_{\rm hot \, shell\,
with\, flat \,space} =
\nonumber\\
&\frac12\int_0^{r_+}b(r)dr- \pi r_+^2\,.
\label{actiondiff1}
\end{align}
So, if $b(r)$ is such that the integral yields a value
less than $\pi {r}_{+}^2$ then
$I_{\rm hot \, shell\, with \,bh}
<
I_{\rm
hot \, shell\, with\, flat \,space}
$, 
and a tiny hot shell with a black hole inside
is a phase favored to a tiny
hot shell with nothing, or
hot flat space inside.
If $b(r)$ is such that the integral yields a value
greater than $\pi {r}_{+}^2$ then
$
I_{\rm
hot \, shell\, with\, flat \,space}
<
I_{\rm hot \, shell\, with \,bh}
$, 
and a
hot shell with nothing, i.e., a  hot shell with
hot flat space inside
is a phase favored to
a
hot shell with a black hole inside.
For instance, 
for $b(r)=2\pi r$, then $\frac12\int_0^{{r}_{+}}b(r)dr=\frac12 \pi
{r}_{+}^2$. So,
$
I_{\rm hot \, shell\, with \,bh}-
I_{\rm
hot \, shell\, with\, flat \,space}
=-\frac12\pi
{r}_{+}^2$ and the hot shell with a
black hole inside is favored.  Quite general, but not generic, if
$b(r)<4\pi r$ at each $r$, then a hot shell with a black hole inside
is preferable, if $b(r)>4\pi r$ at each $r$ than a hot shell
with hot flat space inside is preferable The case $b(r)=4\pi r$, for
which the two phases coexist, has been  analyzed
in Sec.~\ref{specificeos}.  In general,
for each given $b(r)$, which can be a complicated function, one has to
perform the definite integral
that appears in Eq.~\eqref{actiondiff1} to verify
which phase dominates. There will be a radius $r_+$,
with a definite
value that depends on the parameters appearing on the equation of
state $b(r)$ itself that will separate the dominance of one phase
over the other, i.e., above that $r_+$ one phase dominates,
below it the other phase dominates, and at that $r_+$
both phases coexist.

\subsubsection{Pure black hole versus
hot thin shell: Dominant thermodynamic phases
\vskip -0.2cm}

Let us continue by calculating the difference between the action of a
pure black hole with horizon radius $\tilde{r}_+$ and the action of
a hot thin shell with gravitational radius $\tilde{r}_+$.  Here, it
is worth mentioning again that the boundary data for the ensemble is
comprised by $T$ and $R$, together with $\alpha$, the radius of the
shell.  A pure black hole has no $\alpha$, but one can always think of
the existence of some shell at radius $\alpha$ with zero mass, and so
a pure black hole solution can be thought as part of the ensemble, as
it was the case of $b(\tilde{r}_+)=4\pi \tilde{r}_+$ that we have
seen previously.  We assume that this is the case for other forms of
$b(\tilde{r}_+)$, i.e, a pure black hole solution is part of the
ensemble, and proceed accordingly.

We want to study the case where $I(\tilde{r}_+)\leq 0$, so the
$\tilde{r}_+$ solution under consideration is stable and dominates
over vacuum hot flat space.  The pure black hole action has no matter
and so $I_{\rm \, pure\,black \, hole}= \beta
R\left[1-\tilde{k}(R,\tilde{r}_{+{\rm pure\, bh}} )\right]-\pi
\tilde{r}_{+{\rm pure\, bh}}^2$, see Eq.~\eqref{action1}, where
$\tilde{r}_{+{\rm pure\, bh}}$ is the
large pure black hole horizon
radius for the given boundary data $\beta$ and $R$, i.e., it is
of the type of the
pure black hole
$\tilde{r}_{+2}$ that we met before.  In addition, we have two cases
for the hot shell, one is hot shell with black hole inside and the
other is hot shell with hot flat space inside.

For hot shell with black hole inside we have seen that
$I_{\rm hot \, shell\, with \,bh}= \beta R\left[1-\right.$
\break
$\left.\tilde{k} (R,\tilde{r}_{+{\rm hot \,shell}})\right]-\pi r_+^2
-\frac12\int_{r_+}^{\tilde{r}_{+{\rm hot \,shell}}} b(r)\,dr$,
where $\tilde{r}_{+{\rm hot \,shell}}$
is the large hot shell stable gravitational
radii for the given equation of state
$b(\tilde{r}_+)$
and 
the given boundary data $\beta$ and $R$, i.e.,
these radii are the possible analogs of the
hot shell $\tilde{r}_{+2}$ that we met before.
So,
\begin{align}
&I_{\rm \, pure\,bh}
-
I_{\rm hot \, shell\, with \,bh}=
\nonumber
\\
&
 \beta
R\left[1-\tilde{k}(R,\tilde{r}_{+{\rm pure\, bh}}
)\right]-
\beta
R\left[1-\tilde{k}(R,\tilde{r}_{+{\rm hot \,shell}})\right]
+
\nonumber
\\
&
\frac12\int_{r_+}^{\tilde{r}_{+{\rm hot \,shell}}}b(r)\,dr-
\pi \left(\tilde{r}_{+{\rm pure\, bh}}^2- r_+^2\right)\,.
\label{actiondiffa}
\end{align}
Thus, here due to the convoluted equation, one can only ascertain
which phase dominates when a specific reduced inverse
temperature equation of state $b(r)$ is given.
From the specific equation of state
one finds $\tilde{r}_{+{\rm hot \,shell\, with\, bh}}$
and then computes 
$I_{\rm \, pure\,bh}
-
I_{\rm hot \, shell\, with \,bh}$ through Eq.~\eqref{actiondiffa}.
If the difference is negative the pure black hole phase dominates,
if the difference is zero the two phases coexist,
if the difference is positive the hot shell with a black hole
inside phase dominates.

For a hot shell with hot flat space inside we
have seen that
$I_{\rm hot \, shell\, with \,flat\,space}= \beta
R\left[1-\tilde{k}
(R,\tilde{r}_{+{\rm hot \,shell}})\right]
-\frac12\int_{0}^{\tilde{r}_{+{\rm hot \,shell}}}
b(r)\,dr$,
where $\tilde{r}_{+{\rm hot \,shell}}$
is the large hot shell stable possible gravitational
radii for the given equation of state
$b(\tilde{r}_+)$
and 
the given boundary data $\beta$ and $R$, i.e.,
these radii are the
possible analogs of the hot shell
$\tilde{r}_{+2}$ that we met before.
So, 
\begin{align}
&I_{\rm \, pure\,bh}
-
I_{\rm hot \, shell\, with \,flat\, space}=
\nonumber
\\
&
 \beta
R\left[1-\tilde{k}(R,\tilde{r}_{+{\rm pure\, bh}}
)\right]-
\beta
R\left[1-\tilde{k}(R,\tilde{r}_{+{\rm hot \,shell}})\right]
+
\nonumber
\\
&
\frac12\int_{0}^{\tilde{r}_{+{\rm hot \,shell}}}b(r)\,dr-
\pi \tilde{r}_{+{\rm pure\, bh}}^2\,.
\label{actiondiffb}
\end{align}
Thus, here due to the convoluted equation, one can only ascertain
which phase dominates when a specific reduced inverse
temperature equation of state $b(r)$ is given.
From the specific equation of state
one finds $\tilde{r}_{+{\rm hot \,shell}}$
and then computes 
$I_{\rm \, pure\,bh}
-
I_{\rm hot \, shell\, with \,flat\, space}$ through Eq.~\eqref{actiondiffb}.
If the difference is negative the pure black hole phase dominates,
if the difference is zero the two phases coexist,
if the difference is positive the hot shell with flat space 
inside phase dominates.

\subsubsection{\vskip -0.8cm
Summary of the possible thermodynamic phases for generic
equations of state\vskip -0.2cm}

The end result is that to see which phase of the microscopic
gravitational system dominates, one has to verify
Eq.~\eqref{actiondiff1}.  If the hot thin shell with a black hole
inside dominates, then one has to go Eq.~\eqref{actiondiffa} and see
whether the pure black hole dominates over the hot thin shell or not.
If the hot thin shell with hot flat space inside dominates, then one
has to go Eq.~\eqref{actiondiffb} and see whether the pure black hole
dominates over the hot thin shell or not.

In more detail, to see if a hot shell with
a black hole inside is preferable to a hot shell with nothing inside
and vice versa, one has to perform the calculation from
Eq.~\eqref{actiondiff1} after an equation of state $b(r)$ is given,
and deduce the phase with lower action $I$, i.e., lower free energy
$F$.  Clearly, in particular, if $b(r)<4\pi r$ at each $r$ then a hot
shell with a black hole inside is preferable to a hot shell with
nothing inside, if $b(r)=4\pi r$ both phases coexist, if $b(r)>4\pi r$
at each $r$ then a hot shell with nothing inside is preferable to a
hot shell with a black hole inside.  To see if a pure black hole phase
dominates over a hot thin shell with a black hole inside phase or
dominates over a hot thin shell with flat space inside phase and vice
versa, one has to perform the calculations from
Eq.~\eqref{actiondiffa} or Eq.~\eqref{actiondiffb}, respectively,
after an equation of state $b(r)$ is given and with it the
possible solutions for the gravitational of the
hot shell are found, and then deduce the phase with
lower action $I$, i.e., lower free energy $F$.

\end{document}